\documentclass[useAMS,usenatbib]{mn2e}

\usepackage{amsmath}
\usepackage{comment}
\usepackage{amssymb}
\usepackage[dvipsnames]{xcolor}
\usepackage{graphicx}
\usepackage{graphics}
\usepackage{hyperref}
\hypersetup{
     colorlinks   = true,
     citecolor    = blue
}

\def\apj{{ ApJ}}
\def\apjl{{ApJL}}
\def\apjs{{ ApJS}}

\def\aap{{ A\&A}}
\def\aj{{ AJ}}
\def\mnras{{ MNRAS}}
\def\araa{{ ARA\&A}}

\def\nat {{ Nature}}

\def\pasp{{ PASP}}

\def\prd{{ Phys. Rev. D}}

\def\mr{\mathrm}

\newcommand       \mspy 	{\,{\rm M_\odot \, yr^{-1}}}

\def\rp{r_{\rm p}}
\def\ra{r_{\rm a}}
\def\rTb{r_{\rm T}}
\def\Mh{M_{\rm BH}}
\def\Mb{M_{\rm b}}
\def\ab{a_{\rm b}}
\def\eb{e_{\rm b}}
\def\eTb{e_{\rm Tb}}
\def\rpb{r_{\rm pb}}
\def\rRLO{r_{\rm RLO}}
\def\eRLO{e_{\rm RLO}}
\def\ergs{\rm erg \, s^{-1}}

\def\d{\mr{d}}

\def\mc{\mathcal}
\def\Msun{M_{\odot}}
\def\Rsun{R_{\odot}}
\def\eps{\epsilon}
\def\rg{r_{\rm g}}

\def\OmgK{\Omega_{\rm K}}

\def\cs{c_{\rm s}}

\newcommand{\lrb}[1]{\left({#1}\right)}
\newcommand{\lrsb}[1]{\left[{#1}\right]}
\newcommand{\lara}[1]{\left\langle{#1}\right\rangle}

\newcommand{\myemail}{wenbinlu@berkeley.edu}

\title[Quasi-Periodic Eruptions]{Quasi-periodic Eruptions from Mildly Eccentric Unstable Mass Transfer in Galactic Nuclei}
\author[Lu \& Quataert]
  {Wenbin Lu$^{1, 2}$\thanks{\myemail} and Eliot Quataert$^{2}$\\
  $^1$Departments of Astronomy and Theoretical Astrophysics Center, UC Berkeley, Berkeley, CA 94720, USA\\
  $^2$Department of Astrophysical Sciences, Princeton University, Princeton, NJ 08544, USA}

\begin{document}
\label{firstpage}
\maketitle

\begin{abstract}

We propose that the recently observed quasi-periodic eruptions (QPEs) in galactic nuclei are produced by unstable mass transfer due to Roche lobe overflow of a low-mass main-sequence star in a mildly eccentric ($e \sim 0.5$) orbit.  We argue that the QPE emission is powered by circularization shocks, but not directly by black hole accretion.   Our model predicts the presence of a   time-steady accretion disk that is bolometrically brighter than the time-averaged QPE luminosity, but primarily emits in the extreme-UV.   This is consistent with the quiescent soft X-ray emission detected in between the eruptions in eROSITA QPE1, QPE2, and GSN 069. Such accretion disks have an unusual $\nu L_\nu \propto \nu^{12/7}$ optical spectrum.    The lifetime of the bright QPE phase, $10^2$ -- $10^{3}$ yrs, is set by mass-loss triggered by ram-pressure  interaction between the star and the accretion disk fed by the star itself.  We show that the  stellar orbits needed to explain QPEs can be efficiently created by the Hills breakup of tight stellar binaries provided that (i) the stellar binary orbit is tidally hardened before the breakup due to diffusive growth of the f-mode amplitude, and (ii) the captured star's orbit decays by gravitational wave emission without significant orbital angular momentum diffusion (which is the case for low-mass black holes, $\Mh\lesssim 10^6\Msun$).  We conclude by discussing the implications of our model for hyper-velocity stars, extreme mass ratio inspirals, repeating partial TDEs, and related stellar phenomena in galactic nuclei.   



\end{abstract}


\section{Introduction}\label{sec:intro}


Quasi-Periodic Eruptions (QPEs) are a new class of X-ray sources discovered by  wide-field X-ray surveys \citep{Miniutti2019, giustini20_RXJ1301, Arcodia2021, chakraborty21_XMMJ0249}. These sources typically show hour-long bright X-ray bursts on top of quiescent emission, with a recurrence period of $P\sim 10\rm\, hr$. The spectra are thermal-like in the X-ray band, with higher temperatures $(kT \sim 100\rm\, eV)$ in the eruption phase and lower temperatures $(\sim\! 50\rm\, eV)$ in the quiescent phase. This means that the flux ratio between the eruption and quiescent phases is much larger at higher photon energies --- the flare amplitude reaches a factor of $\sim\!10^2$ near $1\rm\, keV$.  They are found in the nuclei of low-mass galaxies, indicating that the radiation is powered by low-mass supermassive black holes \citep[BHs,][]{miniutti13_GSN069, Miniutti2019,shu17_RXJ1301, shu18_GSN069, Arcodia2021, Wevers2022}.

Many existing models of QPEs are based on mass loss from a star orbiting a BH;  previous works have focused on (i) white dwarf (WD) or He stars in highly eccentric orbits \citep{King2020, King2022, wang22_WD_model, zhao22_He_star}, or (2) main-sequence stars on mildly eccentric or circular orbits \citep{Metzger2022}. We argue in what follows that the first class of models is disfavored both because of mass-transfer instability and because of low rates. We also argue that, in the second class of models, the long viscous timescale disfavors accretion power, and that the flares are powered by shocks instead.

\begin{figure*}
 \centering
\includegraphics[width=0.8\textwidth]{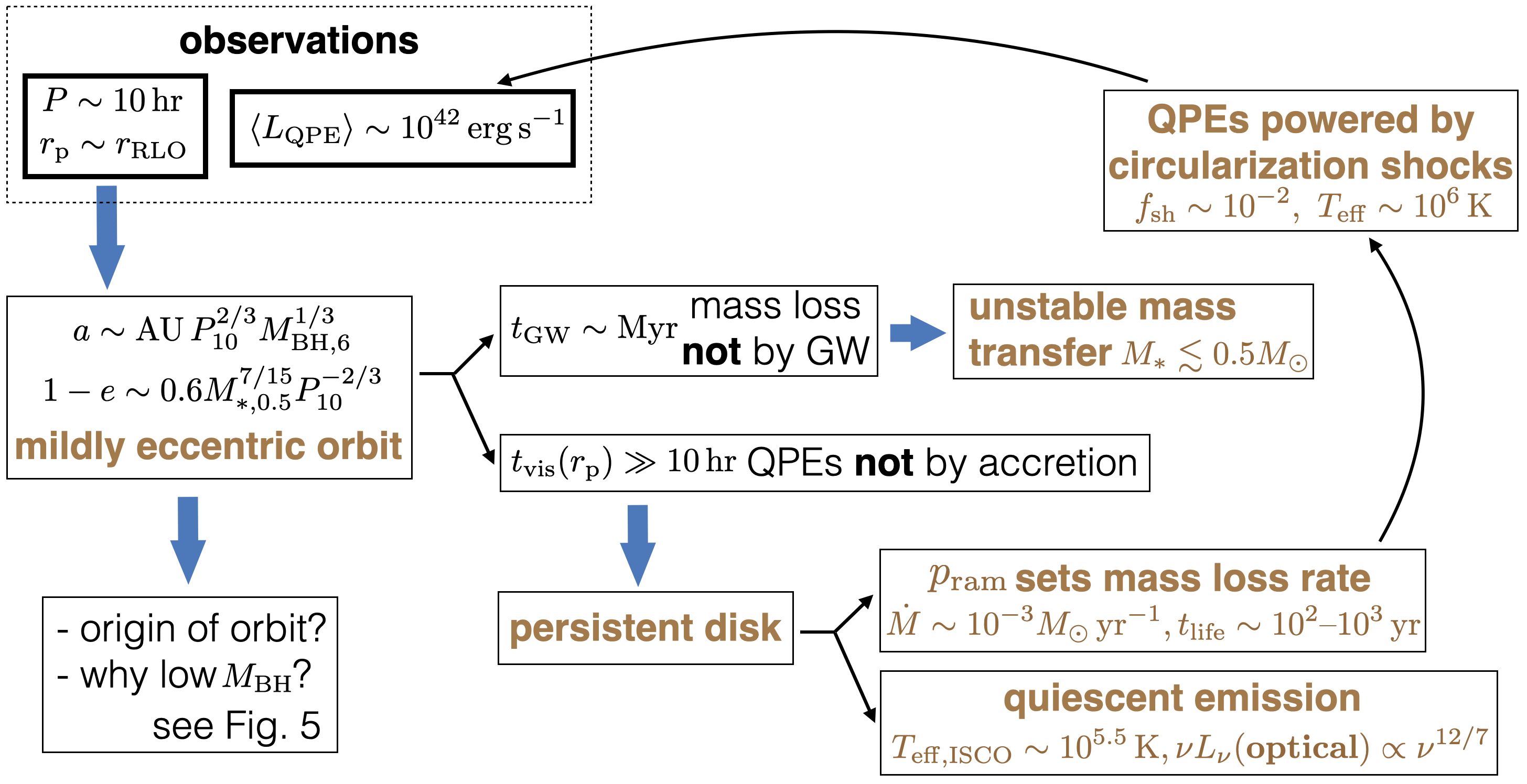}
\caption{Sketch of the model. From the orbital period $P$ and the requirement that the pericenter of the star's orbit $\rp$ is set by the threshold of Roche-lobe overflow $r_{\rm RLO}$, we first derive the semimajor axis $a$ and eccentricity $e$ of the orbit. This then leads to the GW inspiral timescale $t_{\rm GW}$ and a rough estimate of the viscous timescale near pericenter $t_{\rm vis}(\rp)$. The fact that $t_{\rm GW}$ is much longer than the possible lifetime of the QPE sources (as inferred from their luminosity and energy budget) means that the mass-transfer rate is not set by GW orbital decay --- unstable mass-transfer is required (see \S \ref{sec:unstable}). The fact that $t_{\rm vis}(\rp)$ is much longer than the orbital period suggests that QPEs are not directly powered by accretion. Modeling of the quasi-steady disk shows that: (i) the star strongly interacts with the disk gas and the current mass loss rate is set by ram-pressure stripping (\S \ref{sec:sd}), (ii) QPEs can be powered by circularization shocks between the stripped stellar debris and the ambient disk, with radiative efficiencies of perhaps $\sim 1 \%$ (\S \ref{sec:circularization}), and (iii) the quiescent disk emission dominates the time-averaged bolometric luminosity and is mainly in the extreme-UV, but with an unusual optical spectrum of $\nu L_\nu\propto \nu^{12/7}$ (\S \ref{sec:disk}).}
\label{fig:sketch}
\end{figure*}

An alternative class of models are based on possible instabilities in the BH accretion disk \citep{Miniutti2019, sniegowska20_disk_instability, pan22_disk_instability, raj21_disk_tearing}. The simplest version of this, based on thermal-viscous instability \citep{lightman74_thermal_instability}, is disfavored by detailed modeling of the X-ray lightcurves \citep{Arcodia2021}. More complex models \citep[e.g.,][]{pan22_disk_instability, raj21_disk_tearing}, with more free parameters (describing the viscous torques, magnetic pressure, size of the unstable zones, spin-disk misalignment, etc.) are potentially viable for some QPEs that are in bright, long-lived active galactic nuclei (AGN), but these models are inconsistent with the most enigmatic sources (eROSITA QPE1/2) which appear in largely quiescent galaxies \citep{Arcodia2021}.  It is also theoretically unclear why QPE sources would show thermal-viscous instability while the majority of AGN and X-ray binaries do not appear to.

Finally, prior to the discovery of QPEs, a number of authors suggested that interactions between an orbiting star and an accretion disk could power quasi-periodic flares in AGN (e.g., \citealt{Dai2010}).   Aspects of these ideas have subsequently been used to try to explain QPEs (e.g., \citealt{Xian2021,Sukova2021}).   In our model, star-disk interaction plays an important role (\S \ref{sec:sd}) but in ways that are significantly different from previous work.

In this work, we propose a model for QPEs with the following four key ingredients: 
\begin{enumerate}
    \item QPEs are due to Roche-lobe overflow (RLO) of a low-mass main-sequence star;
    \item The star's orbit is only mildly eccentric, with $e \sim 0.5$;
    \item Accretion onto the black hole cannot directly power the QPEs because the viscous time is much longer than the orbital period;
    \item Gravitational-wave (GW) inspiral of the star in its orbit around the BH does not set the mass-transfer rate between the star and the BH in observed QPEs. 
\end{enumerate}
The reasons behind these arguments are presented in \S \ref{sec:prelim} and schematically shown in Fig. \ref{fig:sketch}.

The rest of the paper is organized as follows. In \S \ref{sec:unstable}, we show that the mass transfer from a low-mass main-sequence star (with a convective envelope) to a supermassive BH, initialized by GW-driven orbital decay, will be dynamically unstable, leading to runaway accretion of the star.   Then, as the accretion disk builds in mass, we show in \S \ref{sec:sd} that the mass loss rate from the star is regulated by star-disk interaction in the form of ram-pressure-driven tidal stripping of the star's outer layers. Orbital circularization of stripped stellar debris is  considered in \S \ref{sec:circularization}, where we demonstrate that QPEs can be powered by circularization shocks. In \S \ref{sec:disk}, we model the quasi-steady accretion disk accounting for time-averaged mass and angular momentum source terms and the broad-band disk spectrum is predicted. The origin of the star's current orbit will be addressed in \S \ref{sec:origin}. Other aspects of our model, including predictions for QPE timing variations and related stellar phenomena in galactic nuclei, are discussed in \S \ref{sec:discussion}. A summary and conclusions are presented in \S \ref{sec:summary}.

Near the completion of this work, we became aware of papers by other authors proposing closely related ideas about the origin of QPEs:  \citet{linial_sari23_unstable_mass_transfer} propose that unstable mass transfer by a low-mass star powers QPEs and \citet{Krolik2022} propose that circularization shocks power the observed emission in QPEs.



\section{Preliminaries}
\label{sec:prelim}

We begin by presenting arguments that the four key results listed in \S \ref{sec:intro} are closely related. 

We consider a main-sequence star with a mass of $0.5 M_{*,0.5} M_\odot$ undergoing RLO at pericenter as it orbits a BH of mass $M_{\rm BH} = 10^6 M_{\rm BH,6} M_\odot$ with an orbital period of $P = 10 P_{10}$ hrs. In our model, it is possible for the star to produce two flares per orbit (see \S \ref{sec:inclined_orbits}), so the orbital period may be up to twice the average recurrence period of the fares. This ambiguity does not qualitatively affect our conclusions.

The time dependence of the accretion rate generated by RLO  depends on the eccentricity of the stellar orbit.   For all but the most eccentric orbits, accretion on a viscous time can power a  steady accreting source but cannot produce short duty cycle flares like QPEs.   The reason is that the viscous time of the gas is much longer than the orbital period so that the gas accumulates for many orbital periods before accreting.  Quantitatively, to produce a short duty cycle flare by {accretion} fed by RLO, one requires that the viscous time near pericenter in an eccentric orbit is significantly less than the orbital period.   This in turn requires 
\begin{equation}
1-e \ll 0.03 \, h_{0.1}^{4/3} \, \alpha_{0.1}^{2/3} \ \ \mbox{ (if viscously powered)},
\label{eq:eacc}
\end{equation}
where $h = 0.1 h_{0.1}$ is the dimensionless thickness of the disk and $\alpha = 0.1 \alpha_{0.1}$ is the dimensionless viscosity, both of which set the viscous time of the disk.   This eccentricity constraint is plausibly satisfied in \citet{King2022}'s models of QPE production by white dwarfs in highly eccentric orbits.   Such models face, however, several significant challenges (\S \ref{sec:WD}).  In addition, observations show that there is a persistent disk emitting between eruptions in several of the observed QPEs, inconsistent with the bulk of the disk accreting every orbital period; we return to this in \S \ref{sec:disk}.   In this paper we focus on models of QPEs in which eq. (\ref{eq:eacc}) is {not} satisfied and instead the eccentricity is modest, say $e \sim 1/2$.   We argue in \S \ref{sec:circularization} that the QPE emission is not directly powered by BH accretion, but rather is powered by circularization shocks as the mass lost at pericenter joins the ambient disk. 

We now assess what conditions have to be satisfied for a sub-solar mass star on a mildly eccentric orbit to undergo RLO with orbits compatible with QPE observations.   The requirement is $R_* \gtrsim R_{\rm L} \simeq 0.5a(1-e) (M_*/M_{\rm BH})^{1/3}$.   Using $R_* \simeq R_\odot (M_*/M_\odot)^{0.8}$ for low-mass stars and expressing the semi-major axis $a$ in terms of the orbital period the constraint becomes
\begin{equation}
    (1-e) P_{10}^{2/3} \lesssim 0.6 M_{*,0.5}^{7/15}.
\label{eq:ecc}    
\end{equation}
Equation \ref{eq:ecc} implies stellar orbits with eccentricities of $e \sim 1/2$ for QPE systems if the donor star is indeed a low-mass main-sequence star.   More precisely, any QPE produced by a low mass star with $M_* \lesssim 0.5 M_\odot$ (see \S \ref{sec:unstable}) must satisfy $(1-e)^{3/2} P_{10} \lesssim 0.46$ so in our model QPE systems with orbital period $P \gtrsim 5$ hrs necessarily require stars on mildly eccentric orbits simply to undergo RLO.
In \S \ref{sec:origin} we discuss the origin the stellar orbits through a variant of the widely studied process of the tidal breakup of binaries \citep{hills88_tidal_breakup}.

For the modest eccentricities considered in this paper, the {\em initial} mass-transfer rate as the star first undergoes RLO due to pericenter decay $\dot{r}_{\rm p}$ by GW emission is given by 
\begin{equation}\label{eq:Mdoti}
    \dot M_{\rm ini} \simeq \frac{M_*}{|\rp/\dot r_{\rm p}|} \simeq 2 \times 10^{-7} \ \frac{ M_{BH,6}^{2/3} M_{*,0.5}^2}{P_{10}^{8/3}} \frac{g(e)}{16.8} \, \mspy,
\end{equation}
where $g(e)$ is a function of eccentricity equal to $45, 16.8, 11.3$ for $e = 0.7, 0.5, 0.3$, respectively \citep{Peters1964}.   We argue in \S \ref{sec:unstable} below that eq. (\ref{eq:Mdoti}) does not characterize systems in the observed QPE phase because the mass-loss initiated by GW orbital decay is unstable.    The accretion rates of interest in our problem thus range from eq. (\ref{eq:Mdoti}) to values of $\sim 10^{-3}\mbox{--}10^{-2} \mspy$ that characterize the time-averaged accretion rate in QPE systems as currently observed.   

{In the arguments leading to equations (\ref{eq:ecc}) and (\ref{eq:Mdoti}) we have assumed that the low-mass star has a structure unaffected by its proximity to a massive BH.  This is unlikely to be the case:   tidal heating may modify the structure of the star relative to that of an isolated low-mass star once $\rp$ approaches a few times the tidal disruption radius.  In \S \ref{sec:unstable}-\ref{sec:origin} we neglect the role of tidal heating.   In \S \ref{sec:dynamical_tides} we argue that tidal heating is unlikely to significantly change the conclusions drawn in \S \ref{sec:unstable}-\ref{sec:origin}.}

\section{Unstable Mass Transfer}
\label{sec:unstable}

\begin{figure}
\includegraphics[width = 0.47\textwidth]{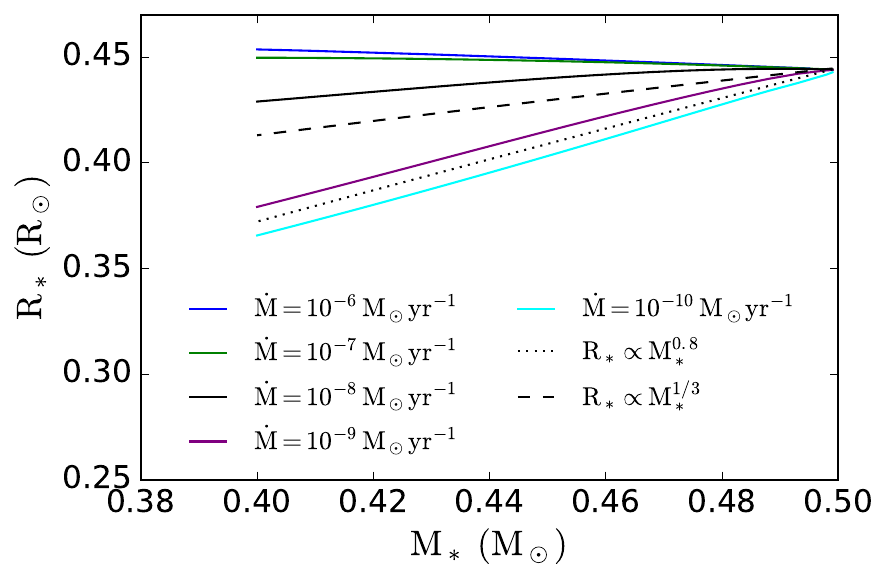}
\caption{Radius evolution of a main-sequence star with initial mass $M_*=0.5\Msun$ under different mass loss rates, as modeled by {\tt MESA}.  For low mass-loss rates, the star maintains thermal equilibrium and $R_*(M_*)$ is similar to that on the low-mass main sequence, $R_* \propto M_*^{0.8}$.   For mass-loss on less than the star's thermal time, however, the radius increases slowly with decreasing mass.  For $R_* \propto M_*^\zeta$ and $\zeta < 1/3$ (dashed line), mass-transfer between a low-mass star and a supermassive BH is unstable if most of the angular momentum of the stripped mass is not returned to the stellar orbit.  We argue in \S \ref{sec:unstable} that this initiates runaway mass transfer leading to the QPE phenomena.}
\label{fig:instability}
\end{figure}

Low-mass stars in which convection dominates the energy transport are prone to expand in radius when they lose mass sufficiently quickly (e.g., \citealt{Ge2015}).    This can lead to runaway unstable mass transfer.  
The initial mass transfer timescale implied by eq. (\ref{eq:Mdoti}) is significantly shorter than the thermal time of a low mass star.   It is thus plausible that the star will expand upon mass transfer.   To assess this, Fig. \ref{fig:instability} shows calculations with {\tt MESA} \citep{Paxton2011,Paxton2013,Paxton2015,Paxton2018,Paxton2019} of the radius of an initially 0.5 $M_\odot$ zero-age main-squence (ZAMS) star as a function of its mass for different mass transfer rates.    For sufficiently slow mass transfer $\dot M \lesssim 10^{-9} \mspy$ the star remains in thermal equilibrium and so the radius scales roughly as $R_* \propto M_*^{0.8}$, as on the lower mass main sequence. By contrast, for the faster mass transfer rates anticipated due to GW inspiral of low mass stars on short-period orbits (eq. \ref{eq:Mdoti}), the stellar radius is roughly constant or expands slightly with decreasing mass.  For comparison, we note that for a fully convective star the adiabatic response to mass loss would be $R_* \propto M_*^{-1/3}$, which is somewhat more rapid expansion than we find for realistic mass-transfer rates in Fig. \ref{fig:instability}.

The stability of mass transfer depends also on how the orbit of the star responds to the loss of mass and orbital angular momentum.    If we define the change in orbital angular momentum as $\dot J/J = \delta (\dot M_*/M_*)$ then $\delta = 0$ implies that no orbital angular momentum is lost (i.e., it is returned to the orbit via gravitational interaction between the star and the disk generated by the mass loss) while $\delta = 1$ corresponds to all of the orbital angular momentum being lost.   For the high mass ratio limit appropriate here, mass transfer is unstable if $\zeta < 2 \delta - 5/3$ where $R \propto M^\zeta$ (e.g., \citealt{Linial2017}).\footnote{\citet{Linial2017} derived this for circular orbits.  The result for eccentric orbits is identical for $\delta = 1$, which we argue is the relevant regime for our problem; for $\delta \ne 1$ the stability condition for eccentric orbits depends in detail on whether the angular momentum is returned to the orbit near pericenter or apocenter.}  Fig. \ref{fig:instability} shows that for the mass-transfer rates expected due to initial RLO, $\zeta \sim 0$ for fully convective stars.  The mass transfer is thus unstable if $\delta \gtrsim 5/6$, i.e., if the majority of the orbital angular momentum associated with the stellar mass-loss is lost from the star's orbit.  This conclusion holds for main sequence stars with masses $\lesssim 0.5 M_\odot$; more massive stars contract under mass loss and their mass-transfer is thus stable even if $\delta \simeq 1$.

There are two arguments that the mass-transfer initiated by low mass (convective) stars will be unstable due to $\delta \sim 1$.   First, given eq. (\ref{eq:Mdoti}), at the very beginning of RLO, {there is a short-lived phase} where the mass-transfer rates are $\sim 10^{-5} \dot M_{\rm Edd}$, {where $\dot M_{\rm Edd}=10L_{\rm Edd}/c^2$ is the Eddington accretion rate of the BH.}  At such low accretion rates there is substantial evidence that accretion proceeds via a hot, geometrically thick, radiatively inefficient accretion flow (RIAF), rather than via a geometrically thin disk \citep{Ho1999,Yuan2014}.   If so, there will be negligible angular momentum exchange between the resulting accretion flow and the orbiting star, leading to unstable mass transfer.  The reason is that the size of the star's Hills sphere $\sim 10^{-2} a M_{\rm BH,6}^{1/3}$ is much less than thickness of the disk and so the vast majority of the disk angular momentum is completely decoupled from the star.

The uncertainty in this conclusion is that the accretion stream generated by RLO is initially very thin so that in principle accretion could proceed via a thin disk at $\dot M \sim 10^{-7} \mspy$.  In this case one might worry that there could be more efficient angular momentum exchange between the star and disk, promoting stability.  However, for the orbital eccentricities $e \sim 1/2 \gg H/r$ considered here, the star will be moving supersonically relative to the gas disk when the star returns to pericenter (we assume rapid circularization of the gas; see \S \ref{sec:circularization}).    The torque between the star and disk for $e \sim 1/2 \gg H/r$ can be modeled as gas drag/dynamical friction (e.g., \citealt{Papaloizou2002,Ida2020}).  Since the star is moving faster than the disk at pericenter, the sign of the interaction is to decelerate the star, transferring its angular momentum to the gas.  This implies that the star is losing additional angular momentum on top of that associated with its stellar mass loss, i.e., $\delta \gtrsim 1$. {The angular momentum lost from the star goes to the outer disk, whose structure will be discussed in \S \ref{sec:disk}.} 

We thus conclude that, independent of some of the uncertainties in accretion physics at very low accretion rates at the onset of RLO, the mass transfer between a fully convective star on an eccentric orbit and a supermassive BH will be unstable; this is true so long as the pericenter distance is sufficiently small that the initial mass transfer generated by GW proceeds on less than the star's thermal time, leading to $\zeta \lesssim 0$ (Fig. \ref{fig:instability}).   Quantitatively, the latter condition can be expressed as $\dot M_{\rm ini} \gtrsim M_*/t_{\rm th,*}$ where $t_{\rm th,*} = GM_*^2/(2 R_* L_*)$ is the star's thermal time.    For low-mass fully convective stars, $L_* \simeq 0.025 \, M_{*,0.5}^2 \, L_\odot$ so that the condition for unstable mass-transfer becomes
\begin{equation}
P \lesssim 12 \, {\rm days} \, M_{\rm BH,6}^{2/3} M_{*,0.5}^{-29/30}.
\label{eq:Punstable}
\end{equation}
In eq. (\ref{eq:Punstable}), we have eliminated the pericenter radius by assuming RLO, i.e., $\rp \simeq 2 R_* (\Mh/M_*)^{1/3}$, and have used the expression for GW pericenter decay appropriate for $\rp \ll a$.   Equation \ref{eq:Punstable} predicts that unstable mass-transfer may lead to QPE-like transients with periods up to $\sim\! 10$ to $100$ days (considering $M_* \simeq 0.1$ to $0.5 M_\odot$), significantly longer than those currently known. {We will discuss the possible origin of such stellar orbits in \S \ref{sec:origin}.}

\section{Stellar Mass Loss Induced by Star-Disk Interaction}
\label{sec:sd}

{In the following we consider how the interaction between the star and the disk that is forming via stellar mass loss at pericenter modifies the structure of the star and ultimately regulates the rate at which tidal stripping by the BH removes mass from the star.   In this section, as well as \S \ref{sec:circularization} and \ref{sec:disk}, we effectively consider the case of a non-spinning black hole, so that there is only one angular momentum axis in the system, that of the stellar orbit.  We return to the generic case of an inclined, eccentric orbit around a rotating black hole in \S \ref{sec:inclined_orbits}.}

As the mass transfer rate from the star increases due to the instability highlighted in \S \ref{sec:unstable}, it will eventually reach $\sim 0.01 \dot M_{\rm Edd} \sim 2 \times 10^{-4} M_{\rm BH,6} \mspy$ at which point radiative cooling causes any hot accretion flow to collapse to a thin disk.    The resulting thin disk is radiation pressure dominated with a dimensionless vertical scale-height
\begin{equation}\label{eq:H_over_R}
    H/r \simeq 5 \times 10^{-3} {\dot M\over 0.01 \dot M_{\rm Edd}} r_{30}^{-1},
\end{equation}
where $r_{30} = r/30 \rg$ is the local radius in the disk (in this section the key radius in the disk will be comparable to the pericenter of the star's orbit).

The dynamical interaction between the star and the disk that it feeds via RLO depends sensitively on the eccentricity of the stellar orbit relative to the dimensionless thickness of the disk $H/r$ \citep{Papaloizou2000}.   This sets whether the star-disk interaction is subsonic ($e \lesssim H/r$) or supersonic ($e \gtrsim H/r$).  Although the high eccentricity case is the one of interest here, we briefly consider the low eccentricity (and inclination) limit $e \lesssim H/r$: in this case, the star resonantly excites spiral density waves that act to repel the gas from the vicinity of the star (i.e., open a gap).  The condition to do so is (e.g., \citealt{Paardekiooper2022})
\begin{equation}
\frac{M_*}{M_{\rm BH}} \gtrsim \left(3 \pi \alpha\right)^{1/2} \left(\frac{H}{r}\right)^{5/2}.
\label{eq:gap}
\end{equation}
In the present problem, using $H/r$ for the radiation dominated disk, as $\dot M$ increases due to runaway mass-transfer, $H/r$ increases and eq. (\ref{eq:gap}) is inevitably violated.  Indeed, even for $\dot M \sim 10^{-2} \dot M_{\rm Edd}$, with $H/r \sim 5 \times 10^{-3}$, efficient gap formation requires $M_* \gtrsim 2 M_{\rm BH,6} M_\odot$, more massive than the stellar masses we are focusing on in this paper. Thus, if the eccentricity were small, the star would not be able to open a gap in the disk in the regime when $\dot M$ is comparable to that inferred in QPE systems. 

In fact, however, the star-disk interaction appropriate for stars on  the eccentric orbits considered here ($e \gg H/r$) is very different from the typical protoplanetary case. The star's orbital velocity at pericenter is $\simeq (1+e)^{1/2}$ times that of the gas and thus the star is moving highly supersonically relative to the gas.  This generates a strong bow shock that mediates the interaction between the star and disk.    Simulations in the protoplanetary context \citep{Duffell2015} show that for eccentricities $e \gtrsim H/r$ even a star that could open a gap by the criterion in eq. (\ref{eq:gap}) would end up shocking against the walls of the gap due to this supersonic relative velocity.

The characteristic ram pressure of the bow shock between the star and disk in QPE sources will be $p_{\rm ram} \sim \rho v^2$ where $\rho$ is the disk density and $v \sim \sqrt{G\Mh /r}$ is the star's orbital speed.  A reverse shock characterized by the same ram pressure will be driven into the star.   If the ram pressure exceeds the photospheric pressure of the star, the shock driven into the star will heat up the outer layers of the star and potentially drive significant mass-loss as the outer layers of the star expand outwards past the Lagrange points.   Our major goal here is to estimate the magnitude of this mass loss and how it depends on the star and disk properties.   The star-disk interaction is concentrated at pericenter because the disk mass and pressure will be largest there and the star's surface layers are tidally stripped by the BH at pericenter.  For the case of misaligned stellar orbit wrt. the disk, the star will interact with the disk twice every orbital period.
The basic picture of ram-pressure-driven RLO, to be described in the following, remains the same. 

To estimate the impact of the disk ram pressure on the structure of the star, we begin with hydrostatic equilibrium for the star, which can be expressed as
\begin{equation}
    \frac{\d p}{\d M_r} = - \frac{G M_r}{4 \pi r^4} \rightarrow \frac{\d p}{\d m_r} = - \bar p \frac{m_r}{\left(r/R_*\right)^4}
    \label{eq:HE}
\end{equation}
where $M_r$ is the enclosed mass and in the second expression we have defined the fractional enclosed mass $m_r = M_r/M_*$ and mean stellar pressure $\bar p = GM_*^2/4 \pi R_*^4$.  Assuming that we focus on a region near the surface of the star with $r \simeq R_*$ and $M_r \simeq M_*$, it is convenient to work with the exterior mass $M_{\rm ex} = M_* - M_r$ rather than the enclosed mass.    Under these near-surface assumptions, eq. (\ref{eq:HE}) can be trivially integrated to yield
\begin{equation}
\frac{p(M_{\rm ex})}{\bar p} \simeq \frac{M_{\rm ex}}{M_*}.
\label{eq:PM}
\end{equation}
That is, the pressure at a depth where the exterior mass is $M_{\rm ex}$ is simply given by $\bar p M_{\rm ex}/M_*$.   

We now consider the star subjected to a time-dependent external pressure $p_{\rm ram}$ due to its interaction with the accretion disk.   If this pressure exceeds the photospheric pressure of the star (which is the case, see below), it will drive a shock into the star to a depth $M_{\rm ex} \simeq M_* p_{\rm ram}/\bar p$.  Neglecting post-shock energy loss (which we assess below), the shocked surface layers of mass $M_{\rm ex}$ may potentially be stripped off from the star near the pericenter of the orbit due to RLO. In Appendix \ref{sec:outer_layers}, we show that adiabatic expansion of the shock-heated layer beyond the Roche lobe leads to a mass loss per orbit $\Delta M_*\sim M_* (p_{\rm ram}/\bar{p})^{\beta}$, where $\beta = 7/5$ (eq. \ref{eq:dMstar_per_orbit}) if the star is marginally filling up the Roche lobe at the pericenter and $\beta=5/3$ (eq. \ref{eq:dMstar_per_orbit_alternative}) if the unperturbed star is far from RLO (i.e., $R_*$ is substantially smaller than $R_{\rm L}$). In the following, we focus on the $\beta=7/5$ case, but our overall conclusions are relatively insensitive to the details of star-disk interactions, which need to be studied in future simulations.
This leads to a time-averaged mass loss rate of the order
\begin{equation}
\lara{\dot{M_*}} \sim \frac{M_*}{P} \lrb{\frac{p_{\rm ram}}{\bar p}}^{7/5},
\label{eq:Mdot-sd}
\end{equation}
where $P$ is the orbital period.

Equation (\ref{eq:Mdot-sd}) is only applicable if the shock penetrates sufficiently deep into the star.  Otherwise, the shocked surface layers can cool sufficiently quickly that the star re-adjusts to thermal equilibrium before the surface layers can expand.  The `quasi-adiabatic' criterion for the applicability of eq. (\ref{eq:Mdot-sd}) is thus that $t_{\rm th} \gg t_{\rm dyn}$ evaluated at the depth $M_{\rm ex} \simeq M_* p_{\rm ram}/\bar p$. This condition corresponds to $\tau(M_{\rm ex}) \gg (c/c_{\rm s})(p_{\rm rad}/p)$ where $\tau(M_{\rm ex})$ is the optical depth down to where the enclosed mass is $M_{\rm ex}$, $c_{\rm s}$ is the sound speed at that depth corresponding to the total pressure, and $p_{\rm rad}$ and $p$ are the radiation and total (gas$+$radiation) pressure. Assuming, as we did in deriving eq. (\ref{eq:PM}) that $r \approx R_*$, it follows that $M_{\rm ex} \simeq 4 \pi R_*^2 \Sigma(M_{\rm ex})$ where $\Sigma(M_{\rm ex})$ is the mass column density of the star to depth $M_{\rm ex}$.   Noting that $\tau(M_{\rm ex}) = \kappa \Sigma(M_{\rm ex})$ the quasi-adiabatic criterion thus becomes
\begin{equation}
\tau(M_{\rm ex}) \simeq \tau_* \frac{p_{\rm ram}}{\bar p} \gg \frac{c}{\cs} {p_{\rm rad}(M_{\rm ex})\over p_{\rm ram}},
\label{eq:ad}
\end{equation}
where $\tau_* = \kappa M_*/4 \pi R_*^2$.
{The structure of the outer layers of the star can be analytically solved under the condition of hydrostatic equilibrium and a polytropic equation of state (see Appendix \ref{sec:outer_layers}). For a $0.5 M_\odot$ star and taking $\kappa \sim 0.4$ cm$^2$/g and polytropic index $\gamma=5/3$, we find that $t_{\rm th}/t_{\rm dyn}\sim 10^6 (p_{\rm ram}/10^{-6}\bar{p})^{3/5}\gg 1$ and hence the adiabatic criterion is well satisfied.}

To apply eq. (\ref{eq:Mdot-sd}) to the accretion disks generated by RLO from the star itself, we use standard estimates of the disk properties as a function of accretion rate, BH mass, etc.   These estimates are good to an order of magnitude, at best, given the significant uncertainties in AGN disk physics, particularly in the radiation pressure dominated regime (e.g., \citealt{Jiang2019}).   Since the derivation of alpha-disk models is standard (e.g., \citealt{Shakura73}) we simply quote our assumptions and results.   We assume that (1) the disk is in steady state with $\dot M \simeq 3 \pi \Sigma \nu_{\rm vis}$ where the viscosity $\nu_{\rm vis} \simeq \alpha c_s^2/\Omega_{\rm K}$ and $c_{\rm s}$ is the total (radiation) sound speed; we return to the validity of the steady state assumption later in the section, (2) radiation diffusion is the dominant energy transport mechanism in the vertical direction, with an opacity given by electron scattering,  (3) the disk obeys the usual relation between effective temperature and accretion rate that is independent of the viscous stress (eq. \ref{eq:diskTeff} below).  
With these assumptions, the disk density and the ram pressure seen by the star are
\begin{eqnarray}\label{eq:pram}
        \rho & \simeq & 2 \times 10^{-8} \, {\rm g \, cm^{-3}} \dot M_{-3}^{-2} M_{\rm BH,6} r_{30}^{3/2} \alpha_{0.1}^{-1}, \nonumber \\
    p_{\rm ram} & \simeq & 10^{12} \, {\rm erg \, cm^{-3}} \dot M_{-3}^{-2} M_{\rm BH,6} r_{30}^{1/2} \alpha_{0.1}^{-1},
\end{eqnarray}
where $\dot{M}_{-3}= \dot{M}/10^{-3}\,M_\odot\rm\,yr^{-1}$.
For comparison, the mean pressure in a star is $\bar p \simeq 3 \times 10^{15} M_{*,0.5}^{-1.2} \, {\rm erg \, cm^{-3}}$ while the photospheric pressure is $p_{\rm ph} \simeq GM_*/(\kappa R_*^2) \simeq 10^5 M_{*,0.5}^{-0.6}\, {\rm erg \, cm^{-3}}$.   Equations (\ref{eq:pram}) thus show that at the small radii near the pericenter distances of the star's orbit, the pressure in a geometrically thin accretion disk is enormous compared to the photospheric pressure of a star and likely within a factor of $\sim 10^5$ of the mean stellar pressure.   Supersonic star-disk interactions at pericenter will thus drive strong shocks into the star that are quasi-adiabatic by eq. (\ref{eq:ad}) and thus drive mass-loss rates of order that estimated in eq. (\ref{eq:Mdot-sd}).


Equations (\ref{eq:pram}) imply that the ram pressure felt by the star in its orbit depends on the accretion rate in the disk.   Equation (\ref{eq:Mdot-sd}) in turn implies that the stellar mass-loss rate depends on the ram pressure felt by the star.   An equilibrium can be reached when $\lara{\dot M_*} = \dot M \equiv \dot M_{\rm eq}$, which yields
\begin{equation}
\dot M_{\rm eq} \simeq {1.6 \times 10^{-3}} \, \mspy \, \frac{M_{\rm BH,6}^{7/19} M_{*,0.5}^{0.7} r_{30}^{7/38}}{P_{10}^{5/19} \alpha_{0.1}^{7/19}}.
\label{eq:Mdoteq}
\end{equation}
This corresponds to a per-orbit mass loss of $\Delta M_* \simeq 2\times10^{-6}\Msun P_{10}^{14/19}$ for our fiducial parameters adopted in the above equation. The equilibrium defined by eq. (\ref{eq:Mdoteq}) appears stable in the sense that the ram pressure $p_{\rm ram} \propto \dot M^{-2}$ (eq. \ref{eq:pram}) and thus the stellar mass loss rate will increase (decrease) in response to a decrease (increase) in the disk accretion  rate (eq. \ref{eq:Mdot-sd}). Of course, the suggested stability in eq. (\ref{eq:Mdoteq}) is misleading because it is based on radiation pressure dominated disk solutions that are known to be thermally and viscously unstable!   Unfortunately, this is the best that we can do theoretically at the present time.


A key assumption of this section is that although the mass transfer between the star and disk is initially dynamically unstable (\S \ref{sec:unstable}) as the disk surface density grows, interaction between the star and the disk it creates determines the structure of the outer layers of the star.   The photosphere is not free to expand as in standard calculations of unstable vs. stable mass transfer (e.g., Fig. \ref{fig:instability}) because the surface is confined by the pressure of the surrounding disk.  We thus argue that the star-disk interaction ultimately leads to a long-lived phase of stellar mass loss at a rate given by eq. (\ref{eq:Mdoteq}).   This predicts quasi-steady mass-transfer rates similar to those inferred observationally in QPE systems, with only a weak dependence on system parameters.   The resulting lifetime for the QPE phase is $\sim 10^2-10^3$ years, set by the stellar mass-loss due to star-disk interaction and tidal stripping.  The viscous time in the disk at $r \sim 30 \rg$ is less than the lifetime of the QPE phase so that the steady-state disk assumption used in eqs. (\ref{eq:pram}) is plausibly satisfied.

\section{Flares Powered by Circularization Shocks}\label{sec:circularization}

\begin{figure}
  \centering
\includegraphics[width = 0.47\textwidth]{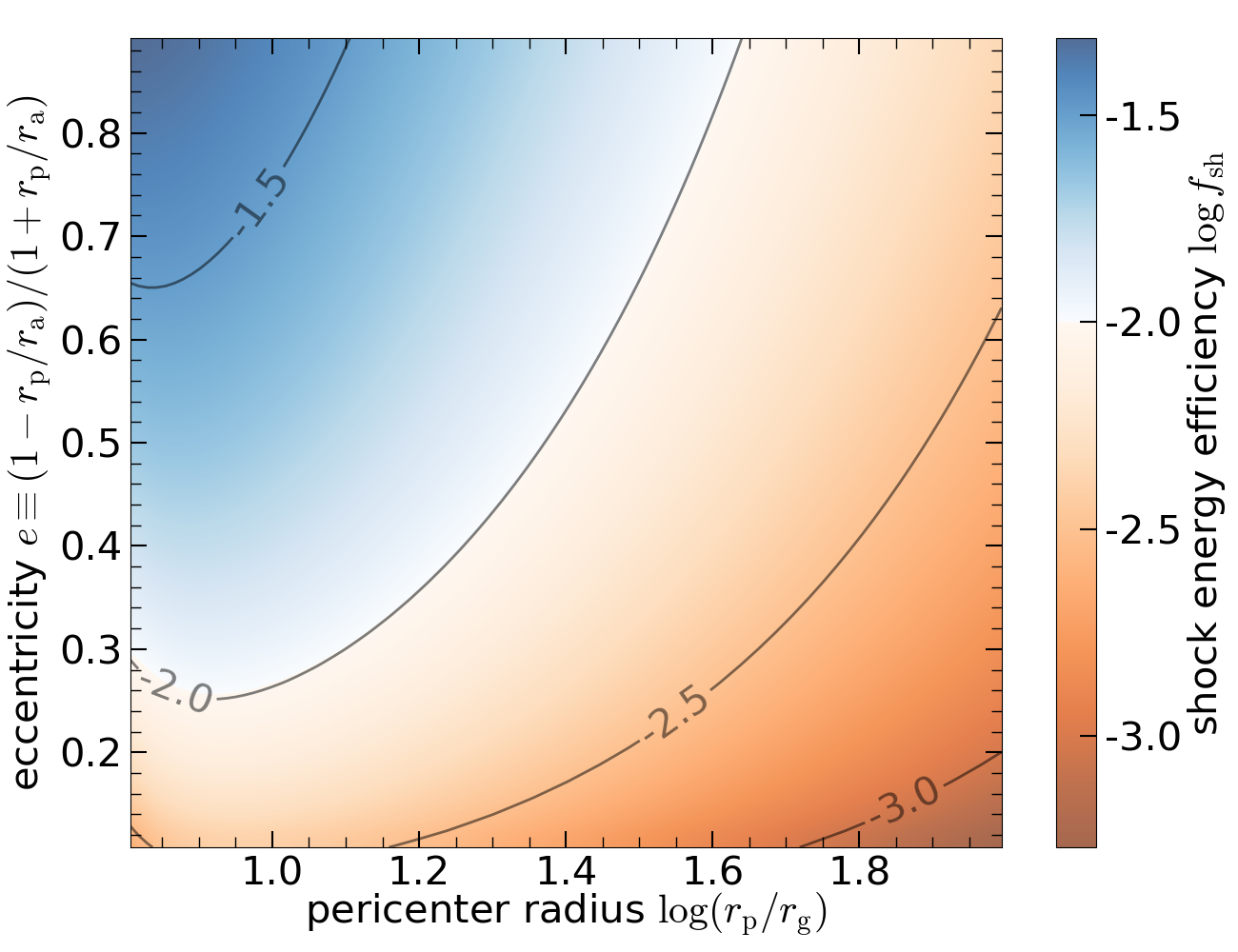}
\caption{The radiative efficiency of circularization shocks for the stripped stellar debris, as given by eq. (\ref{eq:fsh}). The efficiency is define as dissipated energy $= f_{\rm sh} c^2$.  Efficiencies of $f_{\rm sh} \sim 1 \%$ are plausible and can account for QPE energetics, but require that the quiescent disk dominates the time-averaged QPE luminosity (see Fig. \ref{fig:SED}).}
\label{fig:efficiency}
\end{figure}

As argued in \S \ref{sec:prelim}, for orbits that are more circular than eq. (\ref{eq:eacc}), accretion at pericenter cannot directly power a short duty-cycle flare.    Here we argue that such flares can instead be produced as the gas lost at pericenter rapidly circularizes by interacting with the ambient disk.    

Consider a star orbiting around a Schwarzschild BH with pericenter and apocenter radii $\rp$ and $\ra$. The orbital eccentricity is defined as $e\equiv (1-\rp/\ra)/(1+\rp/\ra)$ and the specific orbital angular momentum $J$ and energy $E$ are given by
\begin{eqnarray}
    J^2 = {2\rp^2 c^2\over (\rp/\rg)(1 + \rp/\ra) - 2(1 + \rp/\ra + \rp^2/\ra^2)},\\
    (E/c^2)^2 = \lrb{1 - {2\rg\over \rp}}\lrb{1 + {J^2\over \rp^2c^2}}.
\end{eqnarray}
Each time the star comes back to the pericenter (or runs into the disk in an inclined geometry), the high ram pressure and tidal effects strip a fraction of the star's mass $\Delta M_*/M_*$ as given by eq. (\ref{eq:dMstar_per_orbit}).

This mass loss is initially in the form of a stream leaving from the L1 nozzle (and perhaps the L2 nozzle as well), and then the stream strongly interacts with the disk gas and gets crushed by shocks and fluid instabilities \citep{Klein1994}. This is indeed possible since the total disk mass near $\rp$ is much larger than the per-orbit mass loss from the star $\Delta M_*$, 
\begin{equation}
    M_{\rm disk}(\rp)\sim {\dot{M} \rp^2\over \nu_{\rm vis}(\rp)}.
\end{equation}
Using the scale-height of a radiation-dominated disk as given by eq. (\ref{eq:H_over_R}) and $\dot{M} = \Delta M_*/P$, we obtain
\begin{equation}\label{eq:disk_mass_near_rp}
    {M_{\rm disk}(\rp)\over \Delta M_*} \sim 3\times10^{3} \alpha_{0.1}^{-1} \dot{M}_{-3}^{-2} r_{30}^2 \lrb{1-e\over 0.5}^{-3/2}.
\end{equation}
Note that $M_{\rm disk}(\rp)$ is not the mass of the entire disk, which is dominated by the gas near the outer edge (see \S \ref{sec:disk}).

The stripped debris initially has roughly the same $J$ and $E$ as the star. 
As the debris joins the accretion flow, the radiative efficiency of the circularization shocks, $f_{\rm sh}$, can be estimated by considering that the specific energy of a circular geodesic at the pericenter is $E'/c^2 = (\rp - 2\rg)/\sqrt{\rp^2 - 3\rp\rg}$, so we obtain
\begin{equation}\label{eq:fsh}
    f_{\rm sh} \simeq (E - E')/c^2.
\end{equation}
This is shown in Fig. \ref{fig:efficiency}. 
We find that radiative efficiency of the order $10^{-2}$ is reached for $e\sim 0.5$ and $\rp\sim 30\rg$, which applies to the case of a stellar orbit aligned with the disk plane. The radiative efficiency for the misaligned case, where the relative speed between the disk gas and the star is larger, is generally higher than the co-planar case shown in Fig. \ref{fig:efficiency}, especially for low eccentricities.

In any case, the radiative efficiency of circularization shocks $f_{\rm sh}$ is always lower than the total accretion efficiency of the persistent disk by a factor of a few to 10, so QPEs do not dominate the time-averaged bolometric luminosity of the system (see \S\ref{sec:disk}).   The key to understanding the time-domain manifestation of QPEs lies in the temperatures of the hot gas heated by circularization shocks and the innermost regions of the quiescent disk:  only if the emission from the shock-heated gas is hotter than that from the innermost regions of the disk will there be a large-amplitude X-ray flare at high photon energies. 

{As noted above, the gas stripped from the star initially mixes with the surrounding disk on the ``cloud-crushing timescale" \citep{Klein1994}. However, the stripped gas only fully decelerates and joins the disk, and thus deposits its energy, when it interacts with a comparable amount of mass.  We thus estimate that the emission from the circularization shocks comes from a surface area of the order} 
\begin{equation}\label{eq:hotspot_area}
    A \sim {\Delta M_{\rm *}\over \Sigma(\rp)},
\end{equation}
where $\Sigma(\rp)$ is the disk surface density near $\rp$.


{The duration of the flare, $\Delta t$, depends on the detailed radiation-hydrodynamical interaction between the stripped stellar debris and the disk gas. It is generally in between the dynamical time at pericenter $\Omega_{\rm K}^{-1}(\rp)$ and the diffusion time in the vertical direction $t_{\rm diff}\simeq \rho \kappa H^2/c \simeq (\alpha \OmgK(\rp))^{-1}$ (for a radiation-dominated disk), i.e.
\begin{equation}\label{eq:flaredt}
    \OmgK^{-1} \lesssim \Delta t\lesssim \lrb{\alpha \OmgK}^{-1}\, \rightarrow\, 0.8\mr{\,hr}\lesssim {\Delta t\over  M_{*,0.5}^{0.7}} \lesssim 8\alpha_{0.1}^{-1}\mr{\,hr},
\end{equation}
where 
$\OmgK^{-1}\simeq 0.8\mr{\,hr}\, M_{*,0.5}^{0.7}$ for pericenter radius $\rp\simeq 2R_*(\Mh/M_*)^{1/3}$.
If the transport of radiation is dominated by advection, then $\Delta t$ would be closer to $\OmgK^{-1}$; but if radiative diffusion dominates the energy transport, $\Delta t$ may be closer to $t_{\rm diff}$. The above estimates show that circularization shocks can plausibly explain QPE flare duration of order hours.}

The effective temperature of the flare region is given by
\begin{equation}
    T_{\rm eff,QPE}^4 \simeq {f_{\rm sh} \Sigma(\rp) c^2 \over \sigma \Delta t},
        \label{eq:TeffQPE}
\end{equation}
where $\sigma$ is the Stefan-Boltzmann constant. Using the estimate of the surface density of radiation pressure dominated disks from \S \ref{sec:sd} and taking $\Delta t \sim t_{\rm diff}$, we find
\begin{equation}
    T_{\rm eff,QPE} \simeq 7\times 10^5 \, {\rm K} \ \dot M_{-3}^{-1/4} \, \left(\frac{f_{\rm sh}}{0.01}\right)^{1/4}
    \label{eq:TeffQPE2}
    \end{equation}
where we have assumed electron scattering opacity. 

Equation \ref{eq:TeffQPE2} predicts soft X-ray flares from circularization shocks, reasonably consistent with observations of QPEs; we return to this in \S \ref{sec:disk}.  This estimate of the QPE effective temperature does, however, depend on the uncertain structure of radiation pressure dominated disks (using $\Delta t \sim t_{\rm diff}$ in eq. \ref{eq:TeffQPE} implies $T_{\rm eff,QPE} \propto H^{-1/4}$).    Equation \ref{eq:TeffQPE2} also assumes that photon diffusion is the dominant energy transport mechanism in the vertical direction. Taking the advection limit of $\Delta t \sim \OmgK$ would increase the effective temperature by a factor of $\alpha^{-1/4}\sim 2$.
The color temperature of the radiation is also likely to be somewhat hotter than the effective temperature because the opacity is dominated by electron scattering. {If the last absorption (or thermalization) surface is located at a large scattering optical depth $\tau_{\rm th}$, then the color temperature would be higher than the effective temperature by a factor of the order $\tau_{\rm th}^{1/4}$.} More detailed calculations accounting for this would be useful for better comparison to QPE observations.

Another source of periodic energy dissipation is the bow shock between the star and the disk. The swept-up disk mass by the bow shock is given by $\Delta M_{\rm sw}\sim \pi R_*^2 \rho \Delta \ell$, where $\Delta \ell$ is the path length for each episode of interaction. This should be compared with the per-orbit mass loss from the star $\Delta M_* \sim (p_{\rm ram}/\bar{p})^{\beta}M_*$, and we obtain the following ratio
\begin{equation}\label{eq:swept-up_disk_mass}
    {\Delta M_{\rm sw}\over \Delta M_*} \sim \lrb{ p_{\rm ram} \over \bar{p}}^{1-\beta} \lrb{M_*\over M_{\rm BH}}^{1/3} {\Delta \ell\over r_{\rm p}}.
\end{equation}
For $\beta=7/5$ and our fiducial values of $\bar{p}\sim 3\times10^{15}\rm\, erg\,cm^{-3}$, $p_{\rm ram}\sim 10^{12}\rm\, erg\,cm^{-3}$, and $M_{\rm BH}\sim 10^6\Msun$, we find $\Delta M_{\rm sw}/\Delta M_*\sim 0.3 \Delta \ell/\rp$. We see that the swept-up mass may be comparable to the stellar-mass loss if the interaction length is of the order $\rp$ (appropriate for a stellar orbit embedded in the disk). However, for the misaligned geometry, we expect $\Delta \ell$ to be of the order the vertical scale-height $H$, so we obtain $\Delta M_{\rm sw}/\Delta M_*\sim 10^{-2}$ for accretion rates $\dot{M}\sim 10^{-3}\, \mspy$. Generally, we expect the energy dissipation to be dominated by circularization shocks rather than the bow shock --- for the same reason, the lifetime of a QPE system is set by the mass-loss timescale instead of the timescale associated with angular momentum loss due to the gas drag.

\section{Quasi-steady Accretion Disks in QPE Sources}
\label{sec:disk}
Here we describe aspects of the quasi-steady disk fed by mass loss from the star.  It is important to stress upfront that many aspects of the resulting disk physics are poorly understood, in part due to known difficulties understanding standard AGN spectral energy distribution \citep[SED,][]{Koratkar1999}, as well as the poorly understood physics of radiation-dominated accretion disks more broadly.  As a result, we will focus on highlighting the key differences relative to standard accretion disk models, which may allow the unusual disks envisioned here to be observationally identified.   In the following, we treat the disk emission as a multicolor blackbody --- this assumption is reasonable for the outer disk regions emitting in the UV-optical bands although it likely breaks down for the innermost X-ray emitting regions where the opacity is dominated by electron scattering.

We ignore the complications associated with the circularization of mass from the star and assume that a disk is fed at a time-averaged rate of $\dot M$ at radius $R_0$.    The radius $R_0$ is likely comparable to the Roche radius of the star in its orbit around the BH. \citet{Metzger2012} present analytical solutions for time-dependent viscous disk models given such a delta-function mass source (see their Appendix B1).    The essential physics of such solutions is that most of the mass supplied to the disk accretes onto the central point mass while most of the angular momentum supplied is viscously carried to large radii by a small fraction of the mass.   Thus the disk structure is set by
\begin{eqnarray}
\dot M &=& 3 \pi \nu_{\rm vis} \Sigma = \mr{const},  \ \ \mbox{ for } r < R_0,  \nonumber \\
\dot J &=&  3 \pi \nu_{\rm vis} \Sigma \sqrt{G \Mh r} = \mr{const}, \ \ \mbox{ for } r > R_0.
\label{eq:diskcons}
\end{eqnarray}
The radial surface density profile of the disk thus steepens by $r^{1/2}$ exterior to $R_0$.  

The conservation of angular momentum flux $\dot J$ rather than mass flux $\dot M$ exterior to $R_0$ also changes the local luminosity and effective temperature radiated by each annulus in the disk.  The heating rate per unit area in a viscous disk is $(9/4) \nu_{\rm vis} \Sigma \Omega_{\rm K}^2$ and so the luminosity radiated by a given annulus is $L(r) \propto \nu_{\rm vis} \Sigma G \Mh/r$.    For $\dot M = \rm const$ this implies the well-known result that $L(r) \propto r^{-1}$ and thus $T_{\rm eff} \propto r^{-3/4}$ and $\nu L_\nu \propto \nu^{4/3}$, independent of the form of the viscosity (e.g., \citealt{Balbus1998}).  For radii $r < R_0$, where $\dot M = \rm const$, these results still apply to the disk models considered here.

For radii $r > R_0$, however, $\nu_{\rm vis} \Sigma \propto r^{-1/2}$ (eq. \ref{eq:diskcons}) implies that each annulus in the disk radiates a luminosity $L(r) \propto r^{-3/2}$ and thus $T_{\rm eff} \propto r^{-7/8}$ and $\nu L_\nu \propto \nu^{12/7}$, again independent of the form of the viscosity.    The above considerations for the emissivity of the disk can be summarized with the following effective temperature model
\begin{eqnarray}\label{eq:disk_Teff}
T_{\rm eff}(r) = \left(\frac{3 G \Mh \dot M}{8 \pi \sigma r^3} f(r) \right)^{1/4}, \  \ \ \ (r < R_0)  \nonumber \\
T_{\rm eff}(r) = \left(\frac{3 G \Mh \dot M}{8 \pi \sigma r^3} f(r) \right)^{1/4} \left(\frac{r}{R_0}\right)^{-1/8}, \ \ \ (r > R_0)
\label{eq:diskTeff}
\end{eqnarray}
where the factor $f(r) = 1-(r/r_{\rm ISCO})^{-1/2}$ applies the no torque boundary condition at the innermost stable circular orbit (ISCO, {and we take $r_{\rm ISCO}=6\rg$ as a fiducial value}).  The key prediction relative to standard disk models is thus that the disk spectrum steepens by $12/7-4/3\simeq 0.38$ at wavelengths set by effective temperature corresponding to radius $R_0$.   Specifically, the spectral break due to the change in physics exterior to $R_0$ happens at $h \nu_{\rm break} \simeq 3 k T_{\rm eff}(R_0)$, which corresponds to 
\begin{equation}
h \nu_{\rm break} \simeq 30 \ \mr{eV} \, \dot M_{-3}^{1/4} \, M_{\rm BH,6}^{-1/2} \, r_{30}^{-3/4}. 
\label{eq:nubreak}
\end{equation}

A second feature of the disks fed by RLO at small radii is that there is a finite outer radius of the disk, set by the distance to which the disk can viscously spread during the phase that the star undergoes mass-transfer.   To estimate this radius, we assume that radiation pressure dominates even in the outer disk, which is roughly true for the parameters considered in this paper.   The thickness of the disk for $r \gtrsim R_0$ can then be estimated to be
\begin{equation}
\frac{H}{r} \simeq 15 \, \frac{\dot M}{\dot M_{\rm Edd}} \, \frac{R_0^{1/2} \rg}{r^{3/2}}.
\label{eqHprad}
\end{equation}
This expression for $H/r$ differs from the standard one for radiation pressure dominated disks (e.g., eq. \ref{eq:H_over_R};  \citealt{Shakura73}) by a factor of $(R_0/r)^{1/2} < 1$, because of the weaker dissipation exterior to $R_0$.    The viscous time of the outer disk is thus
\begin{equation}
t_{\rm vis}(r) \simeq 10^4 \, {\rm yr} \, \frac{M_{\rm BH,6}}{\alpha_{0.1}} \left(\frac{\dot M}{\dot M_{\rm Edd}}\right)^{-2} \left(\frac{r}{10^3 r_{\rm g}}\right)^{4.5} \frac{30 r_{\rm g}}{R_0}.
\label{eq:tvis}
\end{equation}
Note that here $\dot{M}$ is the average mass-loss rate from the star but not the local viscous mass accretion rate at $r>R_0$.
For a quasi-steady accretion rate $\dot{M}=\dot{M}_{\rm eq}$ given by eq. (\ref{eq:Mdoteq}) and corresponding source lifetimes of $\sim 10^{2-3}$ yrs (see \S \ref{sec:sd}), an outer edge of the disk of $R_{\rm out} \sim 100 \rg$ is plausible.   If accretion disk winds remove angular momentum from the disk, the outer radius of the disk will be even smaller.

\begin{figure}
\includegraphics[width = 0.47\textwidth]{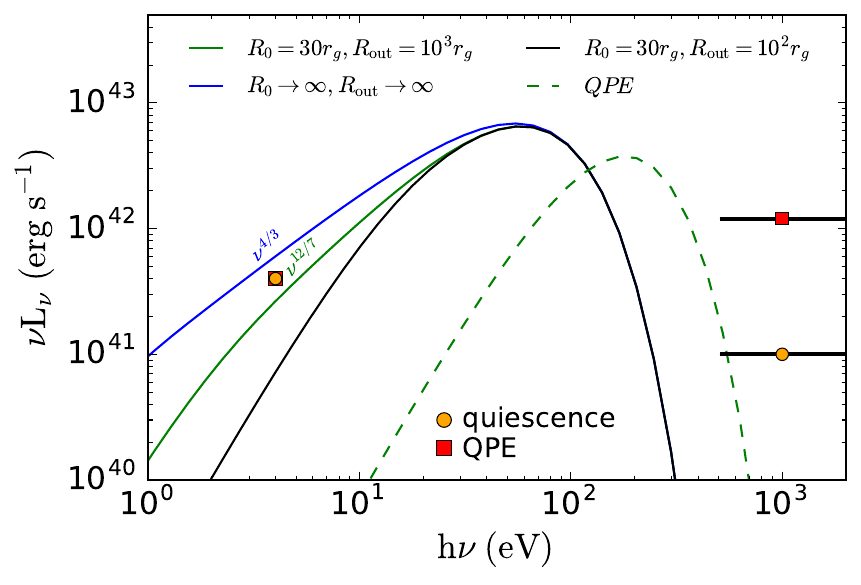}
\caption{Predicted SED of the quiescent time-steady disk (between QPEs; solid lines) and QPE (dashed line; eq. \ref{eq:TeffQPE2}), compared to observations of QPE2 from \citet{Arcodia2021}. During the QPE phase, the total spectrum is the sum of the dashed line and one of the solid lines.  Models asssume $\Mh = 10^6 M_\odot$ and $\dot M = 3 \times 10^{-3} \mspy$. We show quiescent disk SEDs for several $R_0$ (stellar pericenter at which mass is supplied to the disk) and $R_{\rm out}$ (outer radius of the disk due to the finite source lifetime), as well as for a standard multi-color blackbody with $R_0 \rightarrow \infty$ and $R_{\rm out} \rightarrow \infty$.  Our models predict an optical spectrum that is significantly steeper than standard multicolor blackbody spectra ($\nu L_\nu \propto \nu^{12/7}$ instead of $\nu L_\nu \propto \nu^{4/3}$).  We also predict that the QPE is prominent in the soft X-rays but that the quiescent disk dominates the time-averaged bolometric luminosity of the system. }
\label{fig:SED}
\end{figure}

The solid lines in Fig. \ref{fig:SED} show the quiescent (between QPE) SED predicted by eq. (\ref{eq:diskTeff}) for typical QPE parameters in our models:   $\Mh = 10^6 M_\odot$, $\dot M = 3 \times 10^{-3} \mspy$, and a few values of $R_0$ and $R_{\rm out}$.   We also compare to a standard multicolor blackbody with $R_0\rightarrow\infty, R_{\rm out} \rightarrow \infty$.   The key difference is that in disks fed at a finite radius $R_0$, the emission redward of the near-UV is suppressed.  This suppression is by up to a factor of $\sim 3$ in the optical and by even more in the infrared.   

Figure \ref{fig:SED} also compares our predicted SEDs for the quiescent (between QPE) disk to the observations of QPE2 detected by eROSITA \citep{Arcodia2021}.  This source is detected between QPEs using XMM-Newton, consistent with the presence of the quasi-steady accretion disk predicted by our models.  Our effective temperature is slightly lower than the observations, but since the observations are on the Wien tail the flux is very sensitive to small uncertainties in the predicted temperature.  Figure \ref{fig:SED} shows that our quiescent SED models are also consistent with the optical detections of  QPE2 though it is likely that the latter are dominated by stellar continuum and so are upper limits on the accretion disk luminosity; higher spatial resolution observations to resolve out more of the stellar continuum would be valuable.   Finally, the green dashed line in Fig. \ref{fig:SED} shows our predicted QPE spectrum assuming $f_{\rm sh} = 0.01$ and $T_{\rm eff}$ from eq. (\ref{eq:TeffQPE2}).   Particularly given the simplicity of our model, the results are reasonably consistent with the flares in source QPE2 detected by eROSITA.

For eROSITA QPE2 (period $P\simeq 2.4\rm\,hr$), the data are consistent with the quiescent emission dominating the time-averaged bolometric luminosity but the QPE amplitude being large in the X-ray band because the QPE radiation is harder (see \citealt{Arcodia2021}, their extended data Fig 5b).  The same is true for the QPE in GSN 69 (see \citealt{Miniutti2019}, their extended data Figs 5 \& 6).   These trends are consistent with our model in which circularization shocks power the QPE but do not generally dominate the bolometric luminosity of the system. {Our model (eq. \ref{eq:Mdoteq}) predicts that QPEs with longer periods should have fainter quiescent emission in the X-ray band, because of the lower accretion rates and hence lower effective temperatures. } For eROSITA QPE1 ($P\simeq 18.5\rm\, hr$), the quiescent X-ray emission is {indeed relatively weaker}, and the QPE flares dominate the time-averaged bolometric luminosity down to $\sim 0.3$ keV.  Our model also predicts that there should be a relatively bright unobserved UV source in all QPEs as shown in Fig. \ref{fig:SED}.

\section{Origin of the Star}\label{sec:origin}

\begin{figure*}
  \centering
\includegraphics[width = 0.8\textwidth]{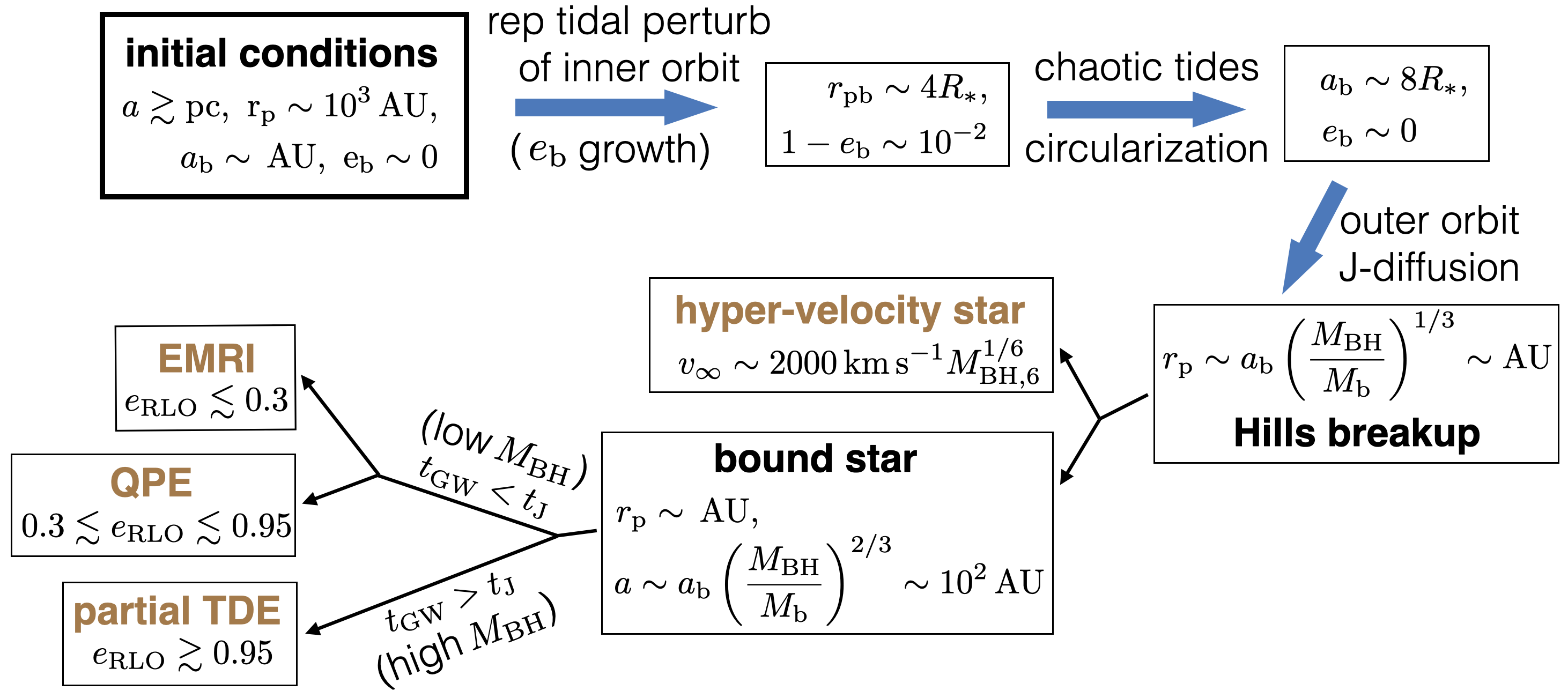}
\caption{Schematic model for the origin of the star's orbit; see \S \ref{sec:origin} for details. The initial conditions are a typical binary with inner SMA $\ab\sim 1\rm\,AU$ in an orbit around the BH with outer SMA $a\gtrsim \rm pc$ (as required by eq. \ref{eq:full_loss_cone_requirement}) and pericenter radius $\rp\sim 10^3\rm\, AU$ (as required by the cumulative eccentricity growth of the inner orbit, $\Delta \eb\sim 1$, due to repetitive tidal perturbations by the BH). When the inner pericenter decreases to $\rpb\sim 4R_*$ due to the growth of inner eccentricity (see Fig. \ref{fig:rpchaotic}), the amplitude of the f-mode grows diffusively and hence the inner SMA quickly shrinks and we assume that the inner orbit circularizes at $\ab\sim 8R_*$. Subsequently, angular momentum diffusion (J-diffusion) of the outer orbit brings the outer pericenter radius to $\rp \sim 1\rm\, AU$ where the inner binary is tidally broken apart by the Hills mechanism. This leads to the ejection of one star at an asymptotic speed of $\sim \! 2000\rm\, km\,s^{-1}$ and the capture of the other star. The bound star's orbit then evolves due to GW emission and J-diffusion. For low BH masses (roughly $M_{\rm BH}\lesssim 10^6\Msun$), the initial SMA of the bound star (roughly $1\rm\, mpc$) is below the drain radius (eq. \ref{eq:drain_limit_radius}) where field star particles are strongly depleted due to mutual scatterings into their loss-cones, so the bound star undergoes GW inspiral (with J-diffusion playing a sub-dominant role) and will produce QPEs or EMRIs, depending on the eccentricity $\eRLO$ when it starts RLO. For high BH masses (roughly $M_{\rm BH}\gtrsim 10^6\Msun$), the initial SMA of the bound star is above the drain radius, so J-diffusion rapidly changes the pericenter of the orbit causing the star to begin RLO at high eccentricities and hence it undergoes repeating partial TDEs.
}\label{fig:origin}
\end{figure*}

Here we describe a channel that can produce the star's current orbit, as schematically shown in Fig. \ref{fig:origin}.   


The initial conditions are a typical stellar binary with inner semimajor axis\footnote{Hereafter, all inner (or outer) orbital parameters are denoted with (without) a subscript $_{\rm b}$.} (SMA) $\ab\sim \rm\, AU$, moderate eccentricity $\eb\sim 0$ (not very close to 0 or 1), and total mass $\Mb$. The barycenter of the inner binary is in an eccentric outer orbit around the BH with SMA $a\sim \rm pc$ and pericenter radius $\rp\sim 10\ab (\Mh/\Mb)^{1/3}\sim 10^3\rm\, AU$. The reason for the choice of $\rp$ will be explained shortly. The outer orbit has specific angular momentum $J \approx \sqrt{2G\Mh \rp}$. For simplicity, we assume the two stars have equal masses $M_* = \Mb/2$ and their radii are given by $R_*\approx \Rsun  (M_*/\Msun)^{0.8}$. The final outcome is not very sensitive to order-unity variations of the above parameters.

The relaxation timescale at a distance $\sim\rm\! pc$ from the BH is denoted as $t_{\rm rel, pc}$. The outer orbital period is $P=2\pi\sqrt{a^3/G\Mh}$. Due to encounters with other massive bodies, the angular momentum undergoes a random walk (hereafter J-diffusion) with mean-squared change per orbit ${\lara{\Delta J^2}} = {P J_{\rm c}^2/t_{\rm rel,pc}}$,
where $J_{\rm c} = \sqrt{GMa}$ is the angular momentum of a circular orbit at the same SMA. 

Due to the tidal gravity of the BH, the inner-orbit eccentricity changes by $\delta \eb$ in each outer orbit as given by \citep[][their eq. 8]{heggie96_inner_eccentricity}
\begin{equation}\label{eq:e_excitation}
    {\delta \eb\over \eb} \simeq \eta^{-3/2} \sqrt{1-\eb^2},
\end{equation}
where we have defined a pericenter penetration factor
\begin{equation}
    \eta = \rp/\rTb,
\end{equation}
and $\rTb$ is the critical pericenter radius for tidal breakup of the binary system \citep{hills88_tidal_breakup}
\begin{equation}\label{eq:tidal_break_up_radius}
    \rTb \simeq \ab\lrb{\Mh\over \Mb}^{1/3}=100\mr{AU}\, a_{\rm b,AU} \lrb{M_{\rm BH,6}\over \Mb/\Msun}^{1/3}.
\end{equation}
Note that eq. (\ref{eq:e_excitation}) is only accurate when $\eta\gtrsim\,$a few such that the inner orbital frequency is much higher than the angular frequency near the pericenter of the outer orbit. Below, we show that this approximation is reasonable for our purpose because the inner eccentricity can be excited to very high values $\eb\approx 1$ when $\eta\sim 10$ in the empty loss-cone regime (so the binary will not be tidally broken up before the eccentricity grows to $\eb\approx 1$).

The number of outer orbits during which the outer orbital angular momentum stays near $J$ is given by
\begin{equation}\label{eq:loss_cone_condition}
\begin{split}
    N &\sim  {J^2\over \lara{\Delta J^2}} = \eta {2\rTb\over a} {t_{\rm rel,pc}\over P}\\
    &\simeq 10^2\eta\, {a_{\rm b,AU} \over a_{\rm pc}^{5/2}} {M_{\rm BH,6}^{5/6} \over (\Mb/\Msun)^{1/3}} {t_{\rm rel,pc}\over 10\mr{\,Gyr}}.
\end{split}
\end{equation}
For an initial value of $\eb$ that is not very close to 0 or 1, over $N$ outer orbits, the inner eccentricity can grow by an amount\footnote{Here, $\Delta \eb$ is calculated in the conservative case based on diffusive growth of eccentricity (i.e., each $\delta e$ may be positive or negative), which occurs if the orbital orientation of the inner orbit changes rapidly. In the linear case (i.e., $\Delta e = N\delta e$), which occurs if the inner orbital orientation stays unchanged, the eccentricity grows even faster.} of $\Delta \eb\simeq \sqrt{N}\delta \eb \simeq 10 \eta^{-1}$ for our fiducial parameters adopted in eq. (\ref{eq:loss_cone_condition}).
For $\eta\sim 10$, the cumulative eccentricity growth would be $\Delta \eb\sim 1$, and then the two stars will undergo strong tidal interactions with each other.

When the inner-orbit eccentricity exceeds a critical value $\eTb$, tidal interactions between the two stars take over the inner orbit evolution. The critical eccentricity is given by the onset of diffusive growth of tidally excited stellar modes \citep{kochanek92_chaotic_tides, mardling95_chaos},
\begin{equation}
\omega_{\rm f}\Delta P_{\rm b}\sim 1,\ \  \mbox{(setting $\eTb$)}
\end{equation}
where $\omega_{\rm f}\approx 1.5\omega_*$ is the frequency of the dominant quadrupolar $\ell=2$ f-mode for a $\gamma=5/3$ polytropic star \citep{lee86_tidal_capture_f-mode}, $\omega_*\equiv \sqrt{GM_*/R_*^3}$, and $\Delta P_{\rm b}$ is the perturbation to the inner orbital period due to per-orbit energy exchange $\Delta E_{\rm b}$ (see the Appendix \ref{sec:Etidal}) between the f-mode and inner orbit ${\Delta P_{\rm b}/ P_{\rm b}} = \lrb{3/2} {\Delta E_{\rm b}/E_{\rm b}}$.

\begin{figure}
\includegraphics[width = 0.46\textwidth]{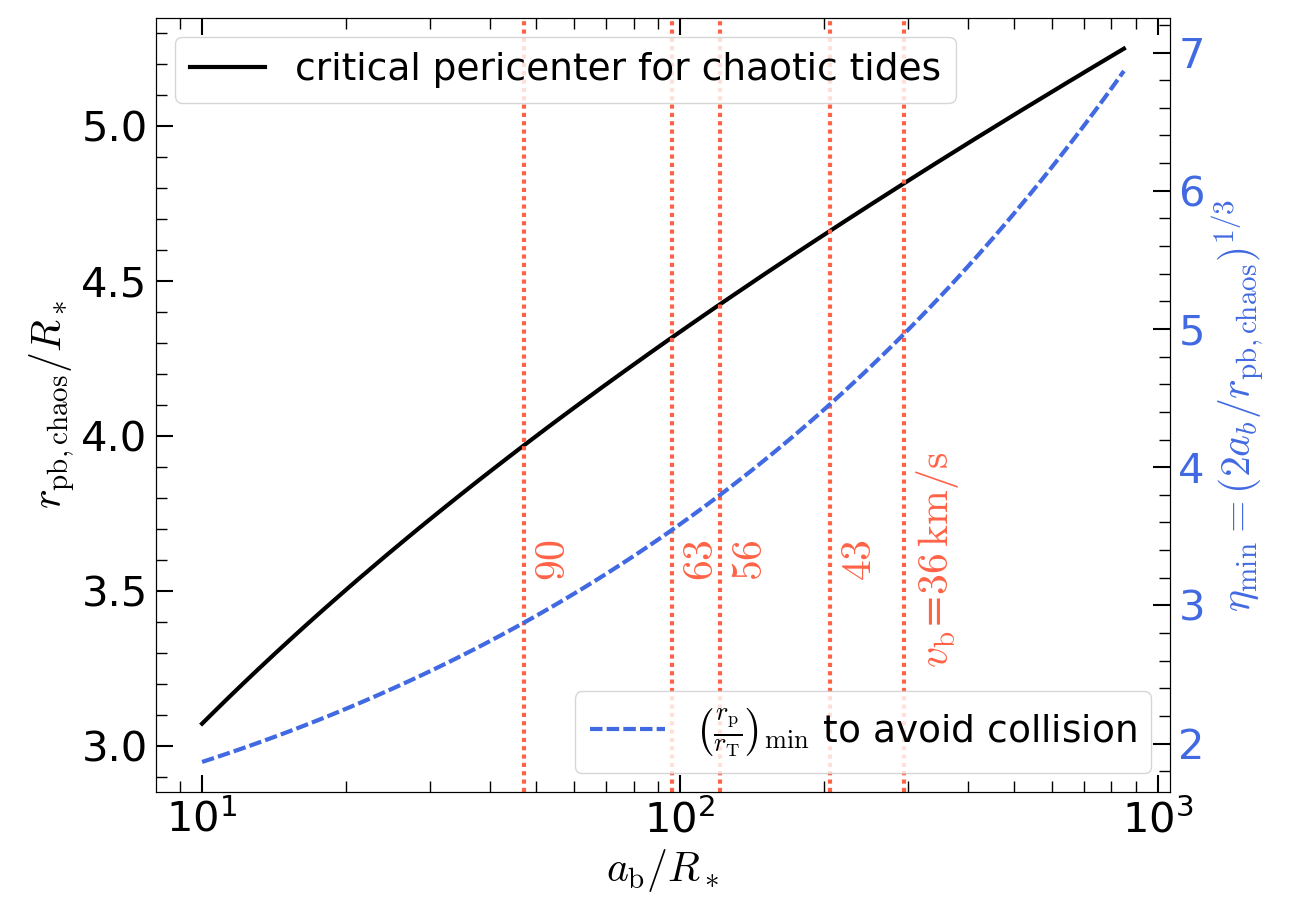}
\caption{The black solid line shows the pericenter radius  $r_{\rm pb, chaos}$ for a stellar binary at which chaotic growth of the f-mode occurs for different initial binary semi-major axes $a_b$; this is for an equal-mass main-sequence binary with $M_*=0.5\Msun$. The blue dashed line shows the minimum ratio $\eta_{\rm min} = (\rp/\rTb)_{\rm min}$ (eq. \ref{eq:eta_min}) to avoid collisions between the two stars. Vertical red dotted lines indicate the inner SMAs corresponding to orbital velocities equal to the host galaxy velocity dispersion for QPE sources measured by \citet{Wevers2022}. Binaries with initial inner SMAs to the right of the red dotted lines tend to evaporate due to encounters with other field stars.}
\label{fig:rpchaotic}
\end{figure}

When $\eb > \eTb$ the amplitude of the f-mode grows diffusively (like a pendulum being kicked at random phases) and the energy grows linearly with the number of inner orbits \citep[e.g.,][]{mardling95_chaos}. This causes rapid orbital circularization, as is likely the case for the orbital evoution of Hot Jupiter planets \citep{wu18_hot_jupiter_migration, vick19_hot_jupiter,yu22_hot_jupiter_migration}. 

The critical inner pericenter radius $r_{\rm pb,chaos}$ for chaotic/diffusive f-mode growth is shown as a black solid line in Fig. \ref{fig:rpchaotic}. To avoid violent collisions between the two stars, we also require $\delta \eb/(1-\eb)|_{\eTb}\lesssim 1$ (i.e., per-orbit change in $\eb$ needs to be sufficiently small), meaning that
\begin{equation}\label{eq:eta_min}
    \eta\gtrsim \eta_{\rm min} =  [2/(1-\eTb)]^{1/3},\ \  \mbox{(avoiding mergers)}
\end{equation}
which is shown as a blue dashed line in Fig. \ref{fig:rpchaotic}. \citet{bradnick17_tidal_mergers} carried out numerical experiments of gravitational interactions between a binary and a BH, and they found that the majority of the binaries end up with mergers. However, they adopted an equilibrium-tide model \citep{hut81_equilibrium_tide, eggleton01_equilibrium_tide} for the energy dissipation in the inner orbit, whereas in our picture, the evolution of the f-mode is in the chaotic regime such that inner orbital energy is dissipated much more rapidly. As long as eq. (\ref{eq:eta_min}) and eq. (\ref{eq:full_loss_cone_requirement}, see later) are satisfied, the two stars will generally not merge in our model.

After the onset of chaotic tidal interactions in the inner binary, the next time the binary comes back to the pericenter of the outer orbit, we assume that the two stars form a compact binary in a circular orbit with
\begin{equation}\label{eq:ab_after_tidal_circularization}
    \ab\simeq 2r_{\rm pb,chaos} \simeq 8R_*,
\end{equation}
although we note that the inner SMA after tidal circularization may be uncertain by a factor of order unity (due to our lack of understanding of the effects of non-linear tidal dissipation). Tidal heating is expected to be concentrated near the outer layers of the stars where the mode amplitudes are the highest. Non-linear tidal dissipation will likely drive mass loss from the binary system and, based on energy conservation, the fractional mass loss is $\lesssim R_*/\ab$, so the stars will survive the heating.
The resulting compact binary will be very resilient to even strong tidal perturbations by the SMBH \citep[and it is more difficult to excite eccentricity for a circular binary,][]{heggie96_inner_eccentricity}. In the subsequent angular momentum evolution of the outer orbit (on a J-diffusion timescale of $(2\rp/a) t_{\rm rel,pc}$), the binary separation may further decrease due to additional tidal dissipation. For simplicity, we assume the binary separation to stay at $\sim\!8R_*$ up to the potential tidal/Hills breakup.

Another key aspect of our model is that, after tidal circularization, since the tidal breakup radius becomes $\rTb\sim 2\rm\, AU$ (eq. \ref{eq:tidal_break_up_radius}), the compact binary system may now be in the full loss-cone regime (corresponding to $N\lesssim 1$ in eq. \ref{eq:loss_cone_condition}, and taking $\eta=1$), provided that
\begin{equation}\label{eq:full_loss_cone_requirement}
    a \gtrsim 1.4\mr{\,pc}\, M_{\rm BH,6}^{1/3} M_{*,0.5}^{0.2} \lrb{t_{\rm rel,pc}\over \mr{10\,Gyr}}^{2/5} \mbox{ (full loss-cone)}.
\end{equation}
Later on, angular momentum diffusion of the outer orbit may bring the outer pericenter to $\rp \sim \rTb\sim 2\rm\, AU$ and the binary is broken apart by the Hills mechanism \citep{hills88_tidal_breakup}. If eq. (\ref{eq:full_loss_cone_requirement}) is violated, eccentricity growth in the inner binary may again be triggered before the Hills breakup and the binary may merge in a significant fraction of the cases \citep{bradnick17_tidal_mergers}.

After the binary breakup, one star is ejected as a hyper-velocity star (HVS) at a typical velocity
\begin{equation}\label{eq:HVS_velocity}
    v_\infty \simeq \sqrt{GM_{\rm BH}\ab \over \rTb^2}\simeq 2000\mr{\,km\,s^{-1}} M_{\rm BH,6}^{1/6} M_{*,0.5}^{-1/15},
\end{equation}
and the other star is captured in a bound orbit with SMA
\begin{equation}\label{eq:SMA_bound_star}
    a\simeq {\rTb^2/\ab} \simeq 2\times10^2\mr{\,AU}\, M_{\rm BH,6}^{2/3}  M_{*,0.5}^{2/15}.
\end{equation}
As a result of unequal binary mass ratios and the orientations of the binary when it gets tidally broken apart, the velocity of the HVS and the SMA of the bound star may fluctuate around the above values in eqs. (\ref{eq:HVS_velocity}, \ref{eq:SMA_bound_star}) by a factor of order unity. The HVS will escape the galaxy in nearly a straight line. In the following, we focus on the fate of the bound star, which has SMA given by eq. (\ref{eq:SMA_bound_star}) and pericenter radius $\rp \lesssim \rTb\simeq 2\mr{\, AU}\, M_{\rm BH,6}^{1/3} M_{*,0.5}^{7/15}$.

\begin{figure}
\includegraphics[width = 0.47\textwidth]{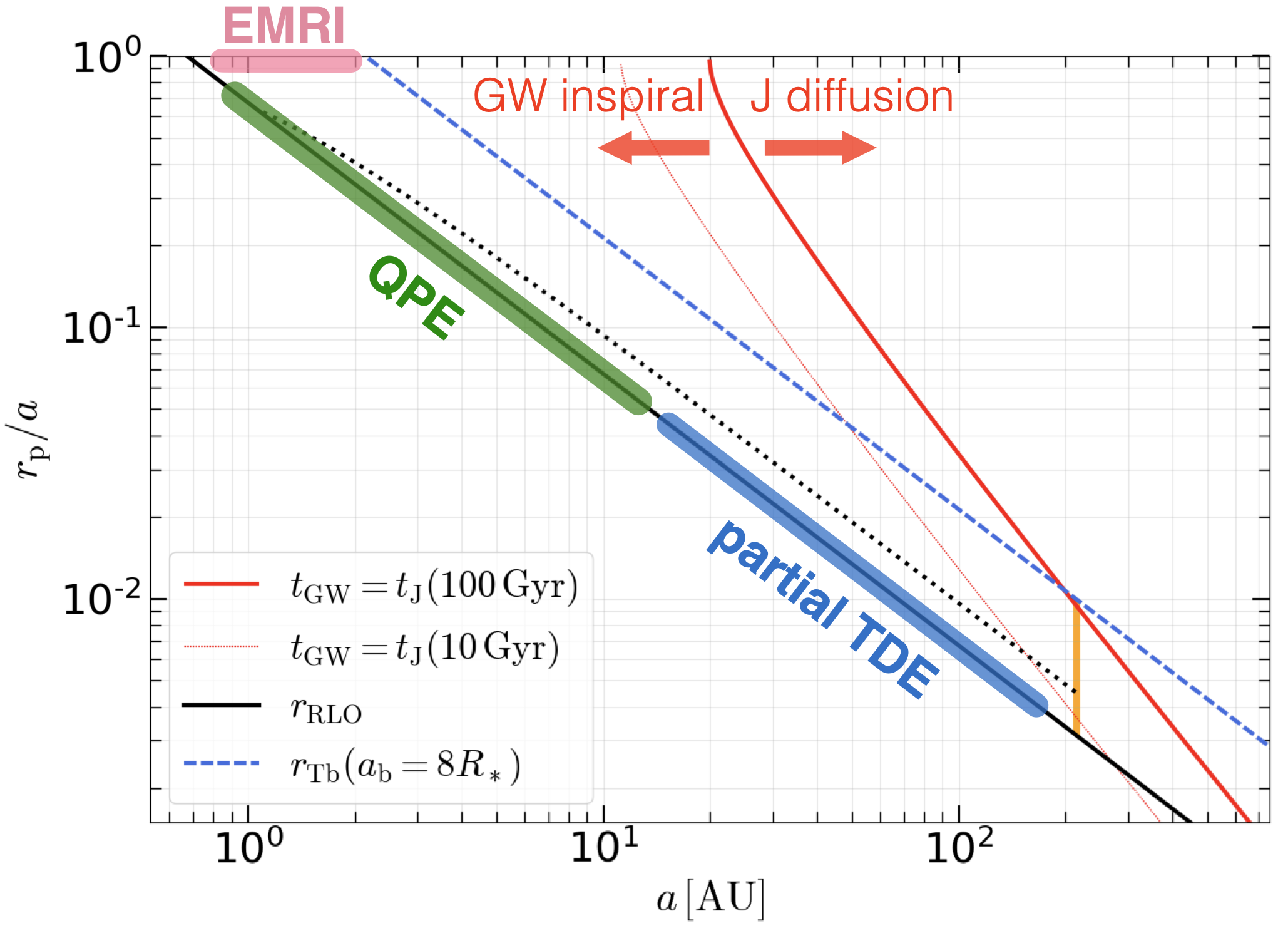}
\caption{Fate of the Hills-captured stars (stellar EMRI, QPE, or repeating partial TDE), under the combined action of GW orbital decay and angular momentum diffusion (J-diffusion).  Note that $\rp = a(1-e)$ so that horizontal lines are lines of constant $1-e$.  Initial orbits for the bound stars have pericenter radius $\rp \lesssim r_{\rm Tb}$ which then evolves to $\rp = r_{\rm RLO}$; the final eccentricity determines the observational manifestation of the resulting stellar Roche-lobe overflow.  The short orange line marks the initial positions of the captured stars. GW emission causes the orbit to decay in both SMA and $\rp$, but much faster in the former; an example of GW-only evolution is shown by the black dotted line. J-diffusion causes the pericenter to stochastically evolve while the semi-major axis is unchanged. Red lines show the positions where the timescale for J-diffusion $t_{\rm J}$ equals to GW inspiral time $t_{\rm GW}$, for two different relaxation times $t_{\rm rel,mpc}=100\rm\, Gyr$ (thick) and $10\rm\, Gyr$ (thin).  Orbital evolution dominated by GW orbital decay preferentially produces QPEs and stellar EMRIs while those dominated by J-diffusion preferentially produce partial TDEs.
}
\label{fig:ae_space}
\end{figure}

The bound star's orbit is subsequently affected by J-diffusion (due to scattering by other stars) and GW emission. Let us denote the relaxation timescale at the SMA of the bound star as $t_{\rm rel,mpc}$ (since its SMA is of the order milli-pc). We fix the SMA of the captured star according to eq. (\ref{eq:SMA_bound_star}) and assume a pericenter distribution as in the full loss-cone regime with
\begin{equation}\label{eq:rp_distribution}
    {\d P\over \d \rp} = {1\over \rTb}, \mbox{ for $\rp < \rTb$}.
\end{equation}
The critical radius for RLO is given by
\begin{equation}\label{eq:RLO_radius}
    \rRLO \simeq 2R_*(\Mh/M_*)^{1/3}.
\end{equation}
We do not model the cases with $\rp < \rRLO$, because the star undergoes either full or partial TDEs in the first pericenter passage after the breakup (this possibility will be discussed in \S \ref{sec:other_phenomena}).

After the Hills capture, we model the J-diffusion process by a Monte Carlo simulation \citep[see][]{lu21_Jdiffusion}. In each orbit, the angular momentum is perturbed by a drift term $\Delta_1J = J_{\rm c}^2P/(2t_{\rm rel, mpc}J)$ and a stochastic term $\Delta_2J = \pm \sqrt{\lara{\Delta J^2}} = \pm (P/t_{\rm rel,mpc})^{1/2} J_{\rm c}$ (random choices between $+$ and $-$), where $J_{\rm c} = \sqrt{GM_{\rm BH}a}$ is the angular momentum of a circular orbit. In the $J/J_{\rm c}\ll 1$ limit, one can geometrically show that the drift in angular momentum is related to the mean squared change by $\Delta_1J = \lara{\Delta J^2}/2J$ \citep{lightman77_J-diffusion}. The drift term tends to circularize a low-J orbit but the stochastic term dominates on timescales shorter than the angular momentum diffusion time $t_{\rm J} = (J/J_{\rm c})^2t_{\rm rel,mpc}$. Meanwhile, we also include orbital decay due to GW emission according to the orbit-averaged angular momentum and energy loss rates given by \citet{Peters1964}. The orbital evolution of the captured star and its final fate is schematically shown in Fig. \ref{fig:ae_space}.

\begin{figure}
\includegraphics[width = 0.47\textwidth]{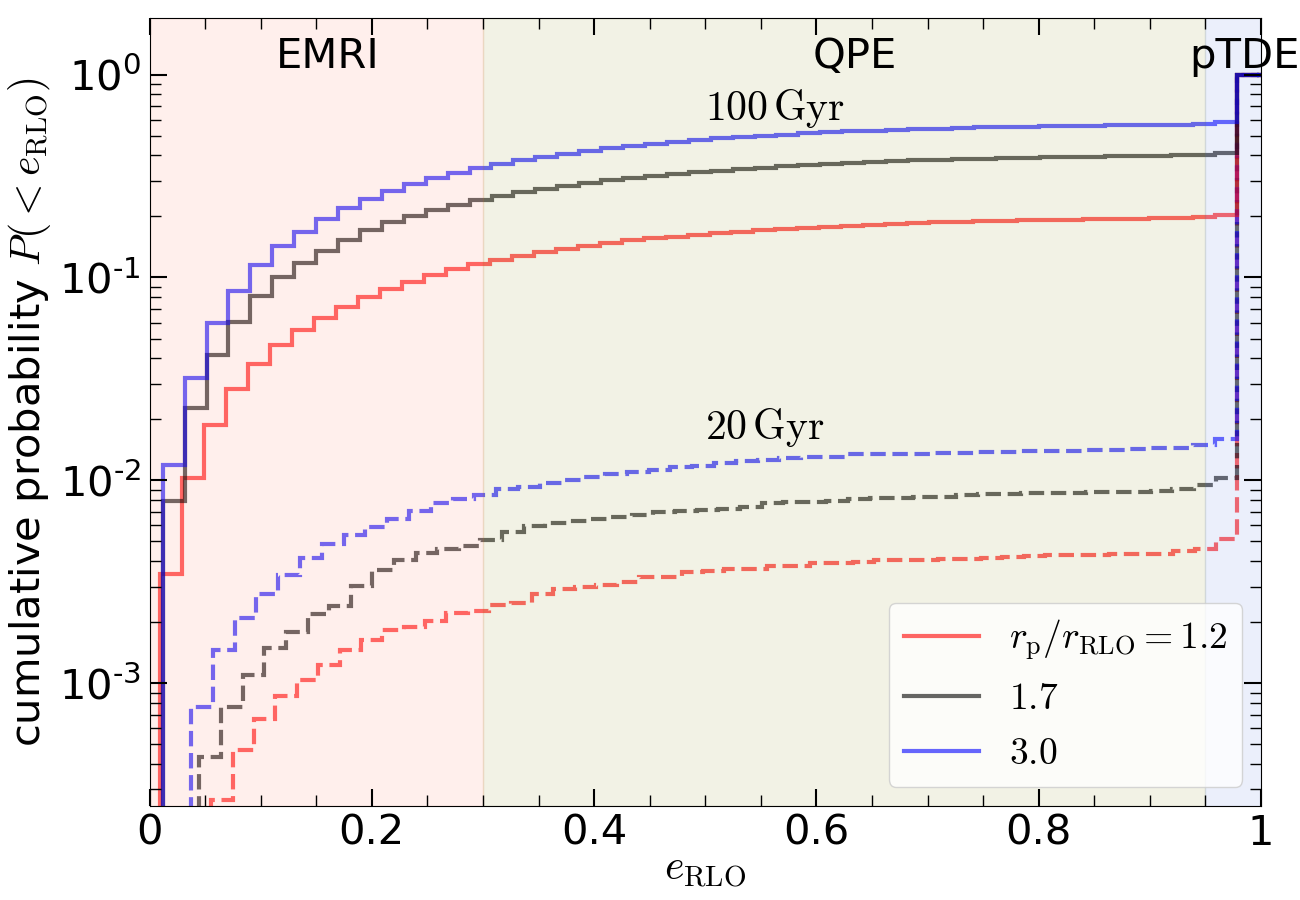}
\caption{The distribution of the eccentricity at the beginning of Roche-lobe overflow, $\eRLO$, for different initial pericenter radii $\rp\in (\rRLO, \rTb)$. Solid lines ($t_{\rm rel,mpc}=100\rm\, Gyr$) and dashed lines ($20\rm\, Gyr$) are for different relaxation timescales. Other parameters in these Monte Carlo calculations are $\Mh=10^6\Msun$, $M_*=0.5\Msun$, $\ab=8R_*$.
}
\label{fig:eRLO}
\end{figure}

We end the Monte Carlo simulation when the pericenter reaches $\rRLO$ and record the eccentricity $\eRLO$ when the star starts RLO. The $\eRLO$ distributions for some of the representative cases are shown in Fig. \ref{fig:eRLO}. We crudely classify these different outcomes based on $\eRLO$: (our conclusions are not sensitive to these exact boundaries)
\begin{enumerate}
    \item Stellar extreme mass-ratio inspirals (EMRIs\footnote{In the literature, the extreme mass-ratio inspiral of a compact object around a supermassive BH is called an EMRI \citep{amaro-seoane18_EMRI}. Here, we (mis)use EMRIs for the inspiral of luminous stars.}) if $\eRLO<0.3$ --- in such a system, the star would undergo mass loss at a roughly constant rate in a nearly circular orbit. In our picture, low-mass stars will still undergo unstable mass transfer in this regime. For a non-spinning BH (or stellar orbits aligned with the disk plane), the unstable RLO will likely produce a source that secularly brightens and fades on relatively long timescales. For a rapidly rotating BH, QPE flares may also be produced by a low eccentricity stellar orbit that is significantly misaligned with the BH spin (see \S \ref{sec:inclined_orbits}).
    \item QPEs if $0.3<\eRLO<0.95$ --- the star in such a system would undergo mass loss at a highly variable rate along each orbit. The gas lost from the star undergoes shock interactions with the existing disk, producing bright X-ray eruptions.
    \item Partial TDEs if $\eRLO>0.95$ --- these systems have orbital periods longer than the viscous timescale of a circular disk near the pericenter (for $\alpha\sim 0.1$ and $H/r\sim 0.1$). The time-dependent flares are thus primarily powered by accretion not circularization shocks.   In addition, direct tidal stripping of the star \citep[e.g.,][]{ryu20_partial_TDEs} is more efficient than the mass loss due to ram-pressure effects (which can only remove a tiny fraction, $(p_{\rm ram}/\bar{p})^{7/5}\lll 1$, of the stellar mass per orbit). In these cases, the effects of star-disk interactions are likely subdominant such that each pericenter passage is simply a new partial disruption\footnote{Since the mass of the existing disk is likely comparable to the newly tidally stripped mass in each orbit, circularization of the stellar debris is likely strongly affected by the interaction between the debris stream and disk gas.} of the star.  
\end{enumerate}

If GW emission dominates the post capture orbital evolution, i.e., in the absence of J-diffusion ($t_{\rm rel,mpc}\rightarrow\infty$), there is a one-to-one map between the initial pericenter radius $\rp$ at Hills capture and $\eRLO$ \citep[cf.][]{Peters1964}.
At our adopted boundaries dividing the three outcomes, we have $\rp/\rRLO=1.56$ and $1.018$ for $\eRLO=0.3$ and $0.95$ respectively. Thus, the fractions of different outcomes are given by (for GW inspiral only) $f_{\rm EMRI} = {(\rTb - 1.56\rRLO)/\rTb} \simeq 51\%$ and $f_{\rm QPE} = {(1.56 - 1.018)\rRLO / \rTb} \simeq 17\%$,
where we have used $\rTb/\rRLO = 4/2^{1/3} = 3.17$ for our choices of $\ab$ (eq. \ref{eq:ab_after_tidal_circularization}) and $\rRLO$ (eq. \ref{eq:RLO_radius}). The fraction of repeating partial TDEs is dominated by the cases with $r_{\rm T,*}<\rp < \rRLO$ (which are not included in our Monte Carlo simulations), and we obtain $f_{\rm pTDE}= (1.018\rRLO-r_{\rm T,*})/\rTb\simeq 16\%$, where $r_{\rm T,*}\simeq R_*(\Mh/M_*)^{1/3}$ is the tidal disruption radius of individual stars. When $\rp < r_{\rm T,*}$, both stars will be disrupted and the fraction of double disruptions (dTDEs) is given by $f_{\rm dTDE}= r_{\rm T,*}/\rTb \simeq 16\%$.

\begin{figure}
\includegraphics[width = 0.47\textwidth]{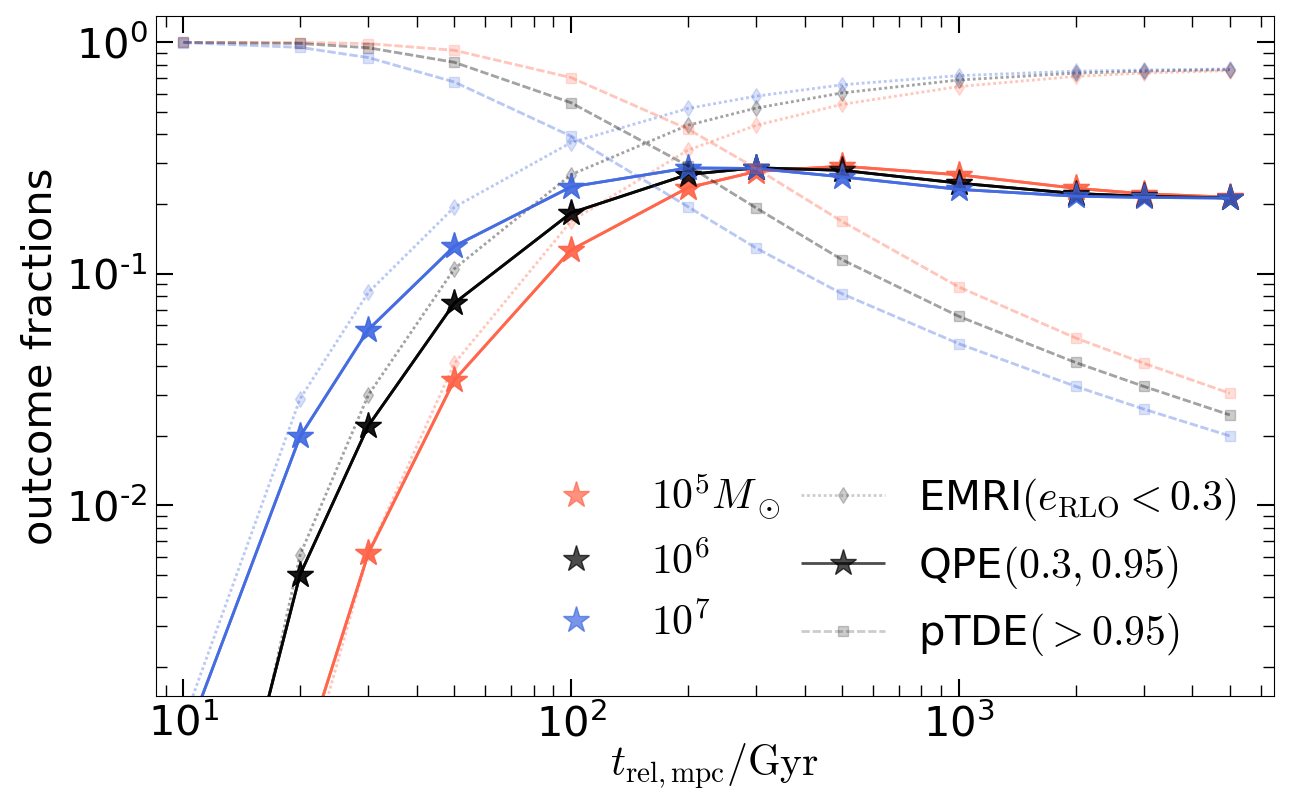}
\caption{Fractions of different outcomes from Monte Carlo simulations of the orbital evolution of the captured star. The initial pericenter of the captured star is sampled according to the full loss-cone probability distribution (eq. \ref{eq:rp_distribution}). Evolution is terminated when the star fills its Roche-lobe near pericenter ($\rp=\rRLO$). The outcome fractions shown in this figure do not include the cases with initial $\rp<\rRLO$, which produce partial or full TDEs (and hence $f_{\rm pTDE}$ shown in this figure is underestimated). Fixed parameters: $M_*=0.5\Msun$, inner SMA before Hills breakup $\ab=8R_*$, and initial SMA of the captured star $a = \ab(\Mh/2M_*)^{1/3}$.
}\label{fig:outcome_fractions}
\end{figure}

In reality, the stellar orbit undergoes both J-diffusion and GW inspiral. Fig. \ref{fig:eRLO} shows that the fractions of the three outcomes strongly depend on $t_{\rm rel,mpc}$, the relaxation time near the SMA of the Hills-captured star.   In Fig. \ref{fig:outcome_fractions}, we show in more detail the dependence of the different outcomes on BH mass and $t_{\rm rel,mpc}$.   The most important parameter is $t_{\rm rel,mpc}$,  with short relaxation times producing primarily partial TDEs and longer relaxation times producing a significant population of EMRIs and QPE.

The relaxation time on mpc scales in galactic nuclei $t_{\rm rel,mpc}$ is a highly uncertain parameter that depends on the poorly understood dynamical processes near a supermassive BH (see, e.g., \citealt{alexander17_stellar_dynamics_nearSMBH}).    A useful constraint on the relaxation time is given by the ``drain limit'' \citep{alexander04_drain_limit} that sets a maximum on the number of star particles at a distance $r\sim \rm mpc$ from the BH:
\begin{equation}\label{eq:drain_limit_number}
    N_{\rm *,max}(r) \sim {0.1 \over \mr{\ln \Lambda}} {M_{\rm BH}^2\over \lara{M_*^2}} {P(r)\over t_{\rm age}},
\end{equation}
where $\ln\Lambda\sim 10$ is the Coulomb logarithm, $\lara{M_*^2}$ is the mean-squared mass of the bodies (depending on their mass spectrum) near radius $r$ from the BH, $P(r)$ is the orbital period for SMA of $r$, and $t_{\rm age}$ is the age of the SMBH (taken to be 10 Gyr in the following). The number of stars near radius $r$ cannot exceed the drain limit, because otherwise their mutual scatterings would drive a large fraction of them into the loss-cone within the age of the system. Solving $N_{\rm *,max}=1$ gives the drain radius $r_{\rm drain}$ below which the stellar density is strongly suppressed by loss-cone depletion
\begin{equation}\label{eq:drain_limit_radius}
    r_{\rm drain} \simeq 100\mr{\,AU}\,  M_{\rm BH,6}^{-1} \lrb{t_{\rm age}\over 10\mr{\,Gyr}}^{2\over 3} \lrb{{\lara{M_*^2}\over 1\Msun^2}}^{2\over 3},
\end{equation}
where we have taken the fiducial value\footnote{\citet{alexander09_strong_mass_seg} proposed that more massive compact objects may dominate the relaxation time in the innermost regions of the stellar density cusp. However, eq. (\ref{eq:drain_limit_radius}) shows that stellar-mass BHs (with masses $\sim\! 10\,\Msun)$ are drained faster such that they are depleted up to a larger radius.} of $\sqrt{\lara{M_*^2}}=1\Msun$.

We compare the SMA of the Hills-captured star (eq. \ref{eq:SMA_bound_star}) and $r_{\rm drain}$ by taking the ratio between them
\begin{equation}\label{eq:drain_limit_comparison}
    {a\over r_{\rm drain}} \simeq 2\, M_{\rm BH,6}^{5/3} \lrb{t_{\rm age}\over 10\mr{\,Gyr}}^{-{2\over 3}} \lrb{{\lara{M_*^2}\over 1\Msun^2}}^{-{2\over 3}}.
\end{equation}
For low-mass supermassive BHs $M_{\rm BH}\lesssim 10^6\Msun$, we find $a\lesssim r_{\rm drain}$ for our fiducial parameters, meaning that the objects that dominate dynamical relaxation are efficiently depleted (with $N_*\lesssim 1$) near the SMA of the Hills-captured star for these galactic nuclei. This means that for the orbital evolution of the Hills-captured stars by these low-mass BHs, J-diffusion is less important as compared to GW inspiral since the relaxation time $t_{\rm rel,mpc}$ is much longer than the Hubble time. However, for high-mass BHs $M_{\rm BH}\gtrsim 10^6\Msun$ (with shorter $t_{\rm rel,mpc}$), J-diffusion plays a dominant role such that most captured stars undergo partial TDEs instead of QPEs or EMRIs.

We now proceed to compare the rate expected for our QPE channel to the observations.   We assume that for $M_{\rm BH}\lesssim10^6\Msun$, $t_{\rm rel,mpc}\gg 10\rm\, Gyr$ because $a\lesssim r_{\rm drain}$.  The rate of QPEs can thus be estimated by
\begin{equation}\label{eq:QPE_rate}
\begin{split}
    \mc{R}_{\rm QPE} &\sim f_{\rm b} f_{\rm full} f_{\rm QPE} {\rTb\over r_{\rm T,*}} \mc{R}_{\rm TDE}\\
    &\sim 10 \mr{\,Gpc^{-3}\,yr^{-1}}\, {f_{\rm b}\over 0.05} {f_{\rm full}\over 0.3} {f_{\rm QPE}\over 0.1} \mc{R}_{\rm TDE,10^3},
\end{split}
\end{equation}
where $f_{\rm b}$ is the binary fraction for the stellar population of low-mass stars near the galactic center, $f_{\rm full}$ is the fraction of the tidally circularized (due to inner eccentricity growth) tight binaries that are in the full loss-cone regime (as required by eq. \ref{eq:full_loss_cone_requirement}), and 
$f_{\rm QPE}$ is the QPE fraction of the bound star after the Hills breakup of the tight binary (as shown in Fig. \ref{fig:outcome_fractions}), $\rTb$ is the critical radius for Hills breakup (eq. \ref{eq:tidal_break_up_radius}), $r_{\rm T,*}=R_*(M_{\rm BH}/M_*)^{1/3}$ is the tidal disruption radius for single stars, and $\mc{R}_{\rm TDE} = 10^3 \mc{R}_{\rm TDE,10^3}\mr{\,Gpc^{-3}\,yr^{-1}}$ \citep{vanVelzen18_TDE_rate} is the rate of TDEs for single stars by these low-mass BHs.

The observed QPE rate can be estimated by the fact that a few sources (e.g., GSN069, eRO-QPE2, RXJ1301) are discovered near redshifts $z\simeq 0.02$ (corresponding to a volume of $V = 4\times10^{-3}\rm\, Gpc^{-3}$). For a typical QPE lifetime of $t_{\rm QPE}$ we have
\begin{equation}\label{eq:obs_QPE_rate}
    \mc{R}_{\rm QPE}^{\rm (obs)} \sim {3\over V t_{\rm QPE}} \sim 7 \mr{\,Gpc^{-3}\,yr^{-1}} \lrb{t_{\rm QPE}\over 10^2\mr{\,yr}}^{-1}.
\end{equation}
The comparison of eqs. (\ref{eq:QPE_rate}) and (\ref{eq:obs_QPE_rate}) shows that our proposed channel is plausible.



\section{Discussion}\label{sec:discussion}

\subsection{Application to Observed QPEs}

Here we briefly discuss the application of our model to GSN 69 \citep{miniutti13_GSN069} and eRO-QPE1 \& QPE2 \citep{Arcodia2021}.    Our goal is to identify the physical parameters (e.g., stellar mass, eccentricity) necessary to roughly account for the observations.  We have already discussed aspects of the quiescent (between flares) emission in these sources in \S \ref{sec:disk} and so do not repeat that analysis here.

\noindent {\bf GSN069:}   The flares in GSN069 have a duration of roughly 1 hr, a period of about 9 hr, and typical peak luminosities of $L_{\rm X} \simeq 5 \times 10^{42} \, \ergs$, corresponding to a time-averaged luminosity of $\langle L_{\rm X} \rangle \simeq 6 \times 10^{41} \, \ergs$.  Taking the recurrence time to be the orbital period, we see that the flares in GSN069 require an orbital eccentricity of $e \gtrsim 0.35$ for $M_* \lesssim 0.5 M_\odot$ (eq. \ref{eq:ecc}) while the duration of the flares may favor a star with a mass somewhat less than $0.5 M_\odot$ (eq. \ref{eq:flaredt}) and a correspondingly larger eccentricity. If the orbital period is twice the reccurrence time, then the constraint on the eccentricity is more stringent, $e\gtrsim 0.6$. The time-averaged luminosity of the source implies a stellar mass-loss rate of $\simeq 10^{-3} (f_{\rm sh}/0.01)^{-1} \mspy$, in line with our theoretical expectations for the mass-loss generated by ram-pressure effects (eq. \ref{eq:Mdoteq}).   There is a rich observed phenomenology for GSN069 \citep{Miniutti2022} that we do not address directly in this work but that would be valuable to study in the context of our model.

\noindent {\bf eROSITA QPE1:}   The flares in eRO-QPE1 have a duration of roughly 7.5 hr, a recurrence time of about 18.5 hr, and time average luminosity of $\langle L_{\rm X} \rangle \simeq  10^{42-43} \ergs$. Assuming $M_*\lesssim 0.5\Msun$, we require an orbital eccentricity of $e \gtrsim 0.6$ (or $\gtrsim 0.75$) if the orbital period is equal to (twice of) the recurrence time (eq. \ref{eq:ecc}). The time-averaged X-ray luminosity corresponds to stellar mass-loss rate of $\simeq 10^{-3}-10^{-2} (f_{\rm sh}/0.01)^{-1} \mspy$, again in line with our theoretical expectations for the mass-loss generated by tidal stripping (eq. \ref{eq:Mdoteq}). The large variation in flare luminosities, as well as occasional overlapping flares \citep{arcodia22_complex_time_evolution}, could arise either from variation in the mass stripped from the star in a given orbit and/or variations in the circularization and radiation processes. Given the complex disk-star interactions (\S \ref{sec:sd}) and stellar debris circularization (\S \ref{sec:circularization}), it is natural to expect such variations.
  
\noindent {\bf eROSITA QPE2:}   The flares in eRO-QPE2 have a duration of order 0.5 hr, a recurrence time of about 2.4 hr, and time-averaged luminosity of $\langle L_{\rm X} \rangle\simeq 3 \times 10^{41} \ergs$.  For this QPE candidate the time between flares is sufficiently short to be consistent with RLO of a star with $M_* \lesssim 0.5 M_\odot$ on a circular orbit (eq. \ref{eq:ecc}).   We favor, though, a lower mass star on a mildly eccentric orbit, e.g., $M_* \simeq 0.1 M_\odot$ and $e \simeq 0.5$ or $M_* \simeq 0.2 M_\odot$ and $e \simeq 0.3$ (or slightly higher eccentricities for an orbital period that is twice the recurrence time).  A lower mass star is also favored by the short flare duration in our model (eq. \ref{eq:flaredt}).  The time-averaged X-ray luminosity in eRO-QPE2 corresponds to a stellar mass-loss rate of $\simeq 3 \times 10^{-4} (f_{\rm sh}/0.01)^{-1} \mspy$, somewhat lower than our default prediction (eq. \ref{eq:Mdoteq}).  We suggest that this is because the radiative efficiency of the circularization shocks $f_{\rm sh}$ is somewhat lower in eRO-QPE2 due to the lower orbital eccentricity.

\subsection{Preference for Low-Mass Host Galaxies}

In the currently small sample of QPEs, a noticeable feature is the preference for low-mass host galaxies --- the hosts of eROSITA QPE1 and QPE2 have stellar masses of $4\times 10^9\Msun$ and $1\times10^9\Msun$ \citep[comparable to that of the Large Magellanic Cloud,][]{Arcodia2021}. This likely indicates that their supermassive BHs have relatively low masses, which provides yet another clue for the origin of QPEs.

A naive guess might be that low-mass BHs have a higher rate at which stars are scattered into the loss cone. However, there is no evidence that this effect is sufficiently strong to explain the host preference of QPEs. For instance, 2-body relaxation (combined with the $M$-$\sigma$ relation) predicts only a weak TDE rate scaling with BH mass $\mc{R}_{\rm TDE}\propto \Mh^{-1/4}$ \citep{merritt13_losscone_dynamics}. Observationally, TDEs do not have a strong preference for low-mass host galaxies \citep{sazonov21_eROSITA_TDE, hammerstein22_ZTF_TDEs}. 

Another possibility is that the Roche radius of a star is slightly smaller for BHs of lower masses $\rRLO\propto M_*^{7/15}\Mh^{1/3}$ so the effective temperature might be higher and hence the emission is easier to detect in the X-ray band. However, this does not produce a sufficiently strong scaling ($T_{\rm eff}\propto \Mh^{-1/6}$ even for a constant luminosity as determined by the flux threshold of the survey), and in fact stellar mass likely plays a more important role than the BH mass.

In our model for the origin of the star producing QPEs, the key requirement is that the orbit of the Hills-captured star decays mainly by GW emission and does not undergo strong J-diffusion. This is only achieved if the relaxation timescale near the SMA of the captured star ($\sim\rm mpc$) is much longer than the Hubble time (see Figs. \ref{fig:ae_space}, \ref{fig:outcome_fractions}). This is seemingly unlikely because the stars near a BH form a cusp with power-law density profile $n_*\propto r^{-\alpha_*}$ with $\alpha_* > 3/2$ such that the relaxation timescale is typically shorter at small radii $t_{\rm rel}(r) \propto r^{\alpha_*-3/2}$ \citep{bahcall76_cusp, alexander09_strong_mass_seg}. 

However, we find that, for sufficiently low-mass BHs ($\Mh\lesssim 10^6\Msun$), the SMA of the Hills-captured star is within the ``drain radius'' $r_{\rm drain}$ of the stellar cusp. The field stars within a distance of $r_{\rm drain}$ deplete themselves by scattering each other into the loss cone \citep{alexander04_drain_limit} --- this effectively creates an empty hole in the innermost region of the cusp where the Hills-captured star undergoes GW inspiral with J-diffusion only playing a minor role. From Fig. \ref{fig:outcome_fractions}, we see that the QPE fraction (as well as the EMRI fraction) strongly depends on the relaxation timescale. This suggests that only low-mass BHs (with longer $t_{\rm rel,mpc}$) would produce QPEs, whereas high-mass BHs (with shorter $t_{\rm rel,mpc}$) would produce repeating partial TDEs instead. This conclusion relies on the still-uncertain stellar dynamics in the central mpc around massive BHs.  Detailed calculations of mass segregation in galactic nuclei would be valuable in testing this hypothesis.

\subsection{Effects of Tides During GW Inspiral}\label{sec:dynamical_tides}

{Dissipation of tidal energy may produce significant heating of the star during GW inspiral prior to RLO.  We neglected this in our analysis of the onset and stability of mass transfer in \S \ref{sec:unstable}.  Here we assess the importance of tidal heating using order of magnitude estimates (see, e.g., \citealt{Generozov2018} for related arguments).   Given the uncertainties (as we shall see), we consider a typical $\sim\, 0.5 M_\odot$ star and keep only the key scaling of tidal effects with stellar  pericenter distance relative to the onset of RLO at $\rRLO \simeq 2 R_* (\Mh/M_*)^{1/3}$.   The total tidal energy is $E_{\rm tide} \sim 10^{-3} G M_*^2/R_* \, (\rp/\rRLO)^{-6}$ i.e.,
\begin{equation}
 E_{\rm tide} \sim 10^{45} \, {\rm erg} \, \left(\frac{\rp}{\rRLO}\right)^{-6}.
 \label{eq:Etide}
\end{equation}   
Most of this energy is stored in the `equilbrium tide' which can be thought of as the stellar f-mode and long wavelength (low-order) sound waves (p-modes).  The linear damping time to dissipate the tidal energy by convective viscosity is $\sim 10^4$ yrs \citep{Kumar1996} and so the heating rate due to linear damping prior to the onset of RLO is at most comparable to the stellar luminosity.}

{For the large tidal amplitudes generated near RLO, the dominant damping process will be nonlinear (e.g., \citealt{Kumar1996,Weinberg2012}).   The efficiency of these nonlinear damping processes is quite uncertain, particularly in convective stars which do not support low frequency internal gravity waves that can resonantly couple to the higher frequency f and p-modes.  \citet{Kumar1996} argue that the dominant nonlinear damping in fully convective stars is non-resonant coupling of the f-mode to higher order sound waves.   Alternatively, the elliptical instability (coupling of the equilibrium tide to inertial waves in a rotating convective object) may be important \citep{LeBars2010,Barker2013}.   
\citet{Kumar1996} estimate the nonlinear damping time of the f-mode to be $\sim 30 \, {\rm days} \, (E_{\rm f}/10^{45} \, {\rm erg})^{-1}$ where $E_{\rm f}$ is the energy in the f-mode.  If we assume that to order of magnitude $E_{\rm f} \sim E_{\rm tide}$, this corresponds to a heating rate $\dot E_{\rm tide} \simeq 3 \times 10^{38} \, {\rm erg \, s^{-1}} (\rp/\rRLO)^{-12}$.   This implies that by $\rp \lesssim 3 \rRLO$ tidal heating may be comparable to or larger than the stellar luminosity.   It is not clear how accurate the \citet{Kumar1996} calculation of the nonlinear decay of a freely oscillating f-mode is for the periodically forced oscillations present in modest eccentricity orbits.  Future calculations are needed to assess this.  \citet{Kumar1996}'s calculation is nonetheless a useful guide to the possible magnitude of the tidal effects.    We now argue that the basic outcome elucidated in \S \ref{sec:unstable} and \S \ref{sec:sd} will remain robust to the uncertainty in the exact radius at which tidal heating becomes important.}

{On general grounds, the non-linear dissipation of tidal energy will  operate most effectively towards the  surface layers of the star, where the dimensionless tidal amplitude is the largest.   Once the tidal heating rate is $\gtrsim L_*$, dissipation of energy in the star will cause the star exterior to where the tidal energy is deposited to expand outwards.   We have verified this explicitly using {\tt MESA} models with additional heating in the near-surface layers, following \citet{Quataert2016}.  The star remains fully convective with roughly constant effective temperature and thus $R_* \propto \dot E_{\rm tide}^{1/2}$ once $\dot E_{\rm tide} \gtrsim L_*$.  Given the strong dependence of tidal heating on $\rp/\rRLO$, tidal heating first becomes dynamically important once $\rp \sim {\rm a \, few \, }\rRLO$ and thus when $\dot E_{\rm tide} \sim {\rm a \, few} \, L_* \ll L_{\rm Edd}$.   At this tidal heating rate convection can easily carry the tidal energy to the surface where it is radiated.    It is also worth noting that the mass-loss driven by tidal heating with $\dot E_{\rm tide} \gtrsim L_*$ has a strict upper bound of $2 \dot E_{\rm tide}/v_{\rm esc}^2 \sim 10^{-9} \mspy \, (\dot E_{\rm tide}/L_*)$.\footnote{This follows from energy conservation:
if the total energy of a putative wind is zero at large radii, the input energy at the base is just that needed to escape the gravity of the star.  Any wind with finite asymptotic energy has a lower mass-loss rate.}    This demonstrates that, if and when tidal heating first reaches $\dot E_{\rm tide} \gtrsim L_*$, the direct mass-loss driven by the tidal heating is negligible for the problem at hand.}

{The above arguments imply that the dominant effect of tidal heating is  to initiate RLO somewhat sooner than would have occurred in the absence of tidal heating, at perhaps $\rp \sim 1-3 \, \rRLO$ depending on the uncertain magnitude of non-linear tidal dissipation.   The star remains fully convective and thus the mass transfer initiated by tidal-heating-induced RLO is unstable by the arguments in \S \ref{sec:unstable}.  Once mass transfer leads to the formation of an accretion disk whose high ram pressure truncates the outer layers of the star (\S \ref{sec:sd}), ram pressure stripping near pericenter will likely be the dominant mass-loss channel, although tidal heating will probably play a role in expanding the outer layers of the star. It is likely that tidal effects in the ram-pressure-confined star are diminished relative to their role in the onset of mass transfer (when the star is not ram-pressure confined by the accretion disk).   The reason is that the outer lower density layers that are most susceptible to non-linear dissipation are absent because of the high ram pressure of the surrounding disk.   More detailed calculations of tidal effects in ram-pressure confined stars would be valuable for better understanding the interplay of ram pressure stripping and tidal heating in the QPE model proposed here.}

\subsubsection{Chaotic Tides Prior to Roche Lobe Overflow}

The tidal perturbations to the stellar eigenmodes may be in the chaotic regime for sufficiently long orbital periods $P\gtrsim 1\mr{\, yr}\, (t_{\rm rel,mpc}/\mr{Gyr})^{1/3}$. This will not be important in the QPE phase but could be earlier in the orbital evolution (and for longer period systems like some partial TDEs).   The dominant contribution to randomizing the phase of the stellar modes relative to the orbit is 2-body scatterings with field objects, which perturb the orbital period by $\Delta P\sim \sqrt{P^3/t_{\rm rel,mpc}}$ in each orbit (the energy exchange between the f-mode and the orbit makes a minor contribution to $\Delta P$). Using a typical f-mode frequency $\omega\simeq 1.5\sqrt{GM_*/R_*^3}$, we obtain $\omega \Delta P\sim 1 (P/\mr{yr})^{3/2} (t_{\rm rel,mpc}/\mr{Gyr})^{-1/2}$ for solar-like stars. When $\omega \Delta P\gtrsim 1$, the f-mode grows in amplitude diffusively and then undergoes non-linear dissipation which heats up the outer layers of the star.
On the other hand, the mode amplitude in stars with orbital periods $P\lesssim 1\mr{\, yr}\, (t_{\rm rel,mpc}/\mr{Gyr})^{1/3}$ stays finite over a large number of orbits, so the energy deposition into the modes is reduced and only potentially becomes important at $\rp \lesssim 1-3 \rRLO$, as discussed earlier in this section.


\subsection{Orbits Inclined Relative to the BH Spin}\label{sec:inclined_orbits}

Throughout this paper we have mainly focused on the case of a 
slowly rotating BH so that there is only one angular momentum axis in the system, that of the stellar orbit.   This is, of course, unlikely to be generically true for stellar RLO in galactic nuclei.  Here we briefly touch on aspects of the more general case of an eccentric stellar orbit inclined relative to the BH spin.

For a rapidly spinning BH, it is possible that the accretion disk near the pericenter radius of the star's orbit ($\rp$) is misaligned with the star's instantaneous orbital plane. Such a misalignment may be due to a combination of two effects: (1) differential Lense-Thirring precession between the nearly circular orbit of the local disk ring near $\rp$ and the star's eccentric orbit; (2) possible \citet{bardeen75_inner_disk_alignment} alignment of the disk near $\rp$ with the BH's equatorial plane.  

The alignment of the gaseous disk due to the \citet{bardeen75_inner_disk_alignment} effect is particularly sensitive to the uncertain structure of radiation-pressure dominated disks.   The reason is that with $H/r \propto 1/r$ (eq. \ref{eq:H_over_R}) the viscous time that governs the dissipation of the warp induced by Lense-Thirring precession is given by \citep{Papaloizou1983} $t_{\rm warp} \sim \alpha (r/H)^2 \Omega_{\rm K}^{-1} \propto r^{3.5}$.  By contrast, the timescale for Lense-Thirring precession of gas on a circular orbit is $\propto r^3$.  Standard radiation-dominated disk scalings thus predict that the entire outer disk is efficiently aligned
(at least out to where gas pressure support begins to dominate)
while the  disk intermediate radii is not necessarily aligned.
Given the sensitivity of this conclusion to mild changes in $H/r$ of radiation dominated disks, we qualitatively consider both limits in which the gas disk near the star's pericenter radius is and isn't aligned with the BH's equatorial plane.

If the gas disk near $\sim \rp$ is not efficiently aligned by the \citet{bardeen75_inner_disk_alignment} effect, it will be roughly aligned with the stellar orbit. 
This is particularly true if the viscous time is less than the Lense-Thirring precession time near $\rp$, so that there is no time for the gas disk to accumulate a significant inclination shift relative to the stellar orbit; this requires $H/r \gtrsim 0.1 (\chi/\pi \alpha)^{1/2} r_{30}^{-3/4}$, where $\chi$ is the dimensionless BH spin (the same conclusion can be reached by comparing the Lens-Thirring precession rate to the rate at which stellar mass-loss supplies angular momentum to the accretion flow).  We note that recent GRMHD simulations of thin, magnetized, tilted disks around a spinning BH show that, even for very thin disks with $H/r\sim 0.015$ to $0.03$, the inner disk is only aligned with the BH spin up to a distance of $\sim\! 10\rg$ \citep{liska19_BPalignment, Liska21_BPalignment}.  This suggests that our analytic estimate of the disk thickness needed for the disk to avoid alignment is quite conservative.   Even if the outer disk does not align with the BH spin, the disk may do so somewhere interior to $\sim \rp$.  This potential warping of the outer disk relative to the inner disk would lead to significant heating of the outer disk by irradiation, which is not included in our estimates of the quiescent disk SED in \S \ref{sec:disk}.  

If the alignment timescale for the gas disk near $\sim \rp$ is short, the resulting disk and QPE dynamics is more complex.  We leave a detailed investigation of this regime to future work, but highlight here a few key points.  The bulk of the disk mass will likely reside near the equatorial plane of the BH, including the gas that viscously spreads to larger radii. The star in a misaligned orbit will undergo two shocks per orbit as it interacts with the ambient disk (neither of which are at pericenter) and the outer layers of the star are perturbed by ram pressure twice per orbit. Continuous mass loss from the star along the orbit means that the stream(s) from the L1 point (and perhaps the L2 point) will interact with the disk twice per orbit. This can potentially produce two flares by circularization shocks. For an eccentric orbit, the mass loss rate due to RLO is the highest near the pericenter, so the flare due to the debris-disk interaction right after the pericenter passage will be brighter than the one preceding the pericenter passage.

An important difference from the star-disk aligned case is that the stellar debris impinges on the disk on an inclined orbit. This significantly reduces the interaction time between the stellar debris and the disk. Based on eq. (\ref{eq:disk_mass_near_rp}), we see that, for our fiducial parameters, the stellar debris can be strongly decelerated by sweeping up a comparable amount of disk mass in a single interaction, which can produce a bright QPE flare. Currently, there are large uncertainties in the hydrodynamic debris-disk interactions, and it is possible that for some parameter space (especially accretion rates much higher than $10^{-3}\,\mspy$) the circularization of the stellar debris might take many orbits, which will not produce bright flares as described in \S \ref{sec:circularization}.

We comment on the regime of a star on a nearly circular orbit that is misaligned with the local disk plane due to the \citet{bardeen75_inner_disk_alignment} effect.  This may occur for some of the cases we have labeled ``EMRI'' in Fig. \ref{fig:outcome_fractions}. {In this case, tidal heating of the star (\S \ref{sec:dynamical_tides}) is much less significant because of the nearly circular orbit.}  The relative velocity between the star and the disk gas is higher at larger inclination angles. We expect both the mass stripped from the star and the radiative efficiency of circularization shocks to increase with the inclination angle --- the highly misaligned cases can potentially produce bright shock-powered X-ray emission at  observed QPE luminosities.  This would blur the distinction we have made between EMRIs and QPEs in Figure \ref{fig:outcome_fractions}.

Aspects of the star-disk interaction in a misaligned geometry  have been considered previously in the QPE context by \citet{Sukova2021, Xian2021}. These authors focused, however, on the interaction between a star and a pre-existing AGN disk.  In our model, the interaction is between a star and the disk that it creates via RLO.
Another key difference is that the star-disk interaction considered by \citet{Sukova2021} is in the regime of very low accretion rates for the pre-existing accretion disk, in which  the ram pressure of the disk material only interacts with the stellar wind and hence does not strip mass away from the body of the star.   This regime is actually difficult to realize for thin accretion disks given the high ram pressures (see eq. \ref{eq:pram}); the neglect of ram pressure is more appropriate for stars interacting with radiatively inefficient accretion flows having $\dot M \ll 0.01\dot M_{\rm Edd}$.   These are, however, unlikely to produce luminous QPEs.

\citet{Xian2021} considered the energy dissipation by the bow shock driven by a star into a misaligned disk, which could potentially produce two flares per orbit. They did not account for the star's mass loss. Since the mass of the disk material swept-up by the star's geometrical cross-section is much less than the mass loss from the star due to tidal stripping (by a factor of $\sim\! 10^{-2}$, see eq. \ref{eq:swept-up_disk_mass}), the luminosity from the circularization shocks driven by the stripped stellar debris is much higher than that from the bow shock. For the same reason, the timescale for the hydrodynamic drags to modify the stellar orbit is longer than the QPE lifetime, which is set by mass-loss rate from the star due to tidal stripping in our model.

\subsection{Potentially Observable General Relativistic Effects}\label{sec:GR_timing}
We suggest that long-term timing of the QPE flares can potentially be used to measure the masses and spins of the underlying BHs.

The star's orbit undergoes general relativistic (GR) apsidal and nodal precessions with per-orbit precessional angles given by \citep[to the lowest order,][]{merritt10_precession_angles}
\begin{equation}
    \phi_{\rm ap}\approx {6\pi \over (1+e)\rp/\rg}, \phi_{\rm nod} \approx {4\pi \chi \over [(1+e)\rp/\rg]^{3/2}},
\end{equation}
where $\chi$ is the dimensionless spin of the BH.
The in-plane apsidal precession produces a shift in the arrival time of QPE flares by an amount
\begin{equation}\label{eq:time_modulation}
    \delta t \simeq {2\rp \sin I \over c} \simeq 0.1\mr{\,hr} {\sin I\over 0.5}\, M_{\rm BH,6}^{1/3} M_{*,0.5}^{7/15},
\end{equation}
at a modulational period
\begin{equation}\label{eq:time_modulation_period}
    P_{\rm ap} = {2\pi P \over \phi_{\rm ap}} = 15\mr{\,d}\, {1+e\over 2} {M_{*, 0.5}^{7/15} P_{10}\over M_{\rm BH,6}^{2/3}},
\end{equation}
where we have taken $\rp\simeq \rRLO$ (eq. \ref{eq:RLO_radius}), $P$ is the stellar orbital period, and $I$ is the angle between the instantaneous orbital angular momentum vector and the line of sight. For $\rp\simeq \rRLO$ and taking the equal sign in eq. (\ref{eq:eacc}), we obtain $(1+e)/2\simeq 1 - 0.3 {M_{*, 0.5}^{7/15}/P_{10}^{2/3}}$.
Lense-Thirring (LT) precession modifies the orientations of both the disk near $\rp$ and the star's orbital plane, and these occur with different periods
\begin{equation}
    P_{\rm LT, *} = {2\pi P \over \phi_{\rm nod}}  = 0.6\mr{\,yr}\, \lrb{1+e\over 2}^{3/2}
    { M_{*,0.5}^{-7/10} P_{10} \over \chi M_{\rm BH,6}},
    \label{eq:LTstar}
\end{equation}
and
\begin{equation}
    P_{\rm LT, disk} = {2\pi \over \phi_{\rm nod}} {2\pi \over \OmgK(\rp)} = 28\mr{\,d}\, {M_{*,0.5}^{7/5} \over \chi M_{\rm BH,6}},
    \label{eq:LTdisk}
\end{equation}
where we have used $e\approx 0$ for the disk.  Due to changes in the projection angle $I$, the flare amplitude and spectrum are expected to be modulated at an order-unity level and the arrival time will also be modulated (eq. \ref{eq:time_modulation}).  

It may be difficult to detect the timing variation in eq. (\ref{eq:time_modulation}) because it is shorter than the duration of each flare. A more promising case is an eccentric stellar orbit that is misaligned with the disk plane. In this case, the disk transit times will be significantly modulated by apsidal precession, similar to geometry of transit timing variations in exoplanets.


As discussed in \S \ref{sec:inclined_orbits}, the outcome of the interaction between the tidally stripped stellar debris with the accretion disk may strongly depend on the relative inclination between the orbital planes of the disk and star. Since the ratio between the two LT precession periods, $P_{\rm LT,*}/P_{\rm LT, disk} = [(1+e)/(1-e)]^{3/2}$, is a factor of a few for mildly eccentric stellar orbits, we expect the observed QPE properties (amplitude, spectrum, etc) to be modulated on a timescale of $P_{\rm LT,disk}$.  It is possible that QPEs will only be efficiently produced for certain inclinations between the star and disk, producing coherent secular variation in the prominence of QPEs, as has been observed in GSN 069 \citep{Miniutti2022}.

If these GR-related timing/amplitude modulations can be observationally identified, this will provide a way of measuring the BH mass and spin in QPEs.

\subsection{Connection to Related Stellar Phenomena in Galactic Nuclei}\label{sec:other_phenomena}

Our model for QPEs predicts that several classes of phenomena should also be produced as by-products. These include: hyper-velocity stars \citep[HVSs,][]{brown15_HVS}, stellar extreme mass-ratio inspirals (EMRIs), repeating partial TDEs, double TDEs (where both stars are tidally disrupted), and stellar merger remnants that are in highly eccentric orbits around the BH. In the following, we roughly estimate the rates of each of these and comment on current or future observations.  We stress that the relative rates of QPEs and other related phenomena depend on the stellar relaxation time in galactic nuclei (Fig. \ref{fig:outcome_fractions}) and how that varies with galaxy properties.  The discussion that follows is based on our current best-guess that the drain limit (eqs. \ref{eq:drain_limit_number} \& \ref{eq:drain_limit_radius}) is the key factor setting the relaxation time at small scales in galactic nuclei, but more work on this complex problem would clearly be valuable.


\subsubsection{Hyper-velocity Stars}

Hills breakup of tight (tidally hardened) binaries give rise to very fast HVSs with typical velocities of $v_\infty\simeq 2000\, M_{\rm BH,6}^{1/6}\rm\,km\,s^{-1}$ (eq. \ref{eq:HVS_velocity}) at a rate that is a few to 10 times higher than the QPE rate. This roughly corresponds to a time-averaged rate of $1 \mr{\,Myr^{-1}\,galaxy^{-1}} \lesssim \mc{R}_{\rm HVS}\lesssim 10\mr{\,Myr^{-1}\,galaxy^{-1}}$ for Milky Way-like galaxies, if $\mc{R}_{\rm QPE}=10\rm\, Gpc^{-3}\,yr^{-1}$ (eq. \ref{eq:obs_QPE_rate}). Note that these HVSs exist in galaxies hosting high- and low-mass BHs, independent of the relaxation time in the innermost regions of the stellar density cusp. A possible example is the recently discovered object S5-HVS1, which was ejected from our Galactic Center about 5 Myr ago at an inferred velocity of $v_\infty\simeq 1800\rm\, km\,s^{-1}$ \citep{koposov20_S5-HVS1}. We note that this population of very fast HVSs was not predicted by the simulations of \citet[][see their Fig. 7]{bradnick17_tidal_mergers}, because they used a different treatment of tidal circularization of the inner binary  (based on the theory of the equilibrium tide) which does not apply when the mode amplitude undergoes diffusive growth. Other works on the velocity distribution of HVSs \citep[e.g.,][]{bromley06_HVS_velocity, rossi14_HVS_velocity, generozov22_HVS_velocity} did not consider the effects of tidal interactions in the inner binary before the Hills breakup.


\subsubsection{Stellar Extreme Mass Ratio Inspirals (EMRIs)}
Our model predicts that galaxies hosting low-mass BHs (roughly $\Mh\lesssim 10^6\Msun$) would produce main-sequence star EMRIs with nearly circular orbits at a rate that is a factor of $\sim\! 3$ higher than the birth rate of QPE sources. For low-mass stars $M_*\lesssim 0.5\Msun$ such that the mass-transfer is unstable, the system produce bright emission in the optical and soft X-ray bands (see Fig. \ref{fig:SED}). They can  potentially be distinguished from other AGN based on their unusual optical spectrum of $\nu L_\nu\propto \nu^{12/7}$ and the fact that they should be bright in the UV but will not have broad-line regions or significant dust IR emission, given the lack of an outer accretion disk and their relatively short lifetimes.   Some of the systems we have labeled ``EMRIs'' may in fact produce QPE-like flares depending on BH spin and disk-BH alignment (see \S \ref{sec:inclined_orbits}).    If the lifetime of these bright EMRIs is comparable to that of QPEs,  the number density of bright EMRIs is of the order $10^3\mr{\,Gpc^{-3}}(\mc{R}_{\rm EMRI}/3\mc{R}_{\rm QPE})$ (cf. eq. \ref{eq:obs_QPE_rate}).

On the other hand, for relatively high-mass stars which undergo stable mass transfer, the accretion rate is much lower, $\dot{M}\sim 10^{-6} M_{\rm BH,6}^{2/3}\mspy$ (for typical stars with $M_*\sim 1\Msun$), which makes them much harder to be electromagnetically detected (especially considering that the accretion will likely be in the RIAF regime with low radiative efficiency). However, these sources are long-lived with a typical lifetime of the order $t_{\rm GW}\sim 10^5\rm\,yr$, and hence their number density is of the order $10^6\mr{\,Gpc^{-3}}(\mc{R}_{\rm EMRI}/10\mr{\, Gpc^{-3}\,yr^{-1}})$. Unfortunately, even for a source at a distance of 10 Mpc, they cannot be detected by future space-based GW detector LISA, mainly because the stars get tidally disrupted at GW frequencies near or below $0.1\rm\, mHz$ before reaching the optimal band (near $1\rm\, mHz$) for LISA sensitivity.  Under our fiducial rate of $\sim\! 10\mr{\, Gpc^{-3}\,yr^{-1}}$, EMRIs provide a minimum accretion rate for low-mass BHs in about 10\% of their cosmic history even when there is no conventional gas-fed AGN; there also should be a number of EMRI-fed weakly active low-mass BHs in relatively nearby ($\lesssim 10\rm\,Mpc$) dwarf galaxies.  If a fraction of these BHs are rapidly rotating, it is likely that the geometrically thick RIAF launches a relativistic jet via the \citet{blandford77_BZ_jet} mechanism. These sources may be unveiled by radio surveys.  

\subsubsection{Repeating Partial TDEs}\label{sec:repeating_pTDE}
In our model, galaxies hosting relatively high-mass BHs (roughly $\Mh\gtrsim 10^6\Msun$) do not produce QPEs or EMRIs efficiently, at least not through the Hills mechanism. This is because J-diffusion of the Hills-captured star's orbit pushes its pericenter radius to the RLO threshold while the eccentricity is still high $1-e\lesssim 10^{-2}$. The star then undergoes periodically repeating partial TDEs, one per pericenter passage. 
These partial TDEs can also occur near lower-mass BHs at a somewhat reduced rate (by a factor of a few to 10) relative to higher-mass BHs, since the pericenter radius of the pre-breakup binary needs to be a bit fine-tuned to be between the tidal disruption radius $r_{\rm T,*}$ and the RLO radius $\rRLO$ for the individual stars (these two radii differ by a factor of $\sim\!2$).

Possible examples of repeating partial TDEs are ASASSN-14ko \citep{payne21_partial_TDE} and eRASSt J0456-20 \citep{liu22_partial_TDE}. ASASSN-14ko is a periodic optical/X-ray transient with $P\approx 114\rm\,d$ near the nucleus of a relatively massive host galaxy with stellar mass $M_*\sim \mr{few}\times10^{10}\Msun$ \citep{payne21_partial_TDE, payne22_14ko_evolution, payne22_14ko_multiband}. eRASSt J0456-20 is a periodic X-ray and UV transient with $P\approx 223\rm\,d$ possibly associated with a high-mass BH with $M_{\rm BH}\sim 10^7\Msun$ based on the velocity dispersion of the host galaxy \citep{liu22_partial_TDE}. In these two cases with relatively high-mass BHs, we indeed expect J-diffusion to play a more important role than GW emission in the orbital decay of the Hills-captured star --- this leads to repeating partial TDEs instead of QPEs or EMRIs.

Partial TDE models for ASASSN-14ko have been considered by many authors \citep{payne21_partial_TDE, curfari22_AS14ko_Hills_capture, liu22_pTDE_AS14ko}. \citet{curfari22_AS14ko_Hills_capture} proposed that ASASSN-14ko's period of 114 days is consistent with the Hills-capture origin (the same argument also applies to eRASSt J0456-20), which is again advocated for here. However, these previous authors did not provide a clear explanation for the tight inner orbit of the binary prior to the Hills breakup, which is needed to reproduce the orbital period of ASASSN-14ko. In our model, the binary is hardened by rapid tidal dissipation in the inner orbit.



\subsubsection{Other Events}
Double TDEs occur when the pericenter radius of the pre-breakup binary is less than the tidal disruption radii of each of the individual stars, $\rp \lesssim r_{\rm T,*}$. Based on our model, the rate of double TDEs is of the same order as the QPE rate, $\mc{R}_{\rm dTDE}\sim 10\rm\, Gpc^{-3}\rm\,yr^{-1}$ (cf. eq. \ref{eq:obs_QPE_rate}). Since both stars are disrupted, there are interesting, unique features such as double-peaked lightcurves \citep{mandel15_double_TDE} and precursor emission due to collisions of the two debris streams \citep{bonnerot19_double_TDE_precursor}. When the TDE sample size grows to several hundred, it is possible that several double TDEs will be discovered. It is also interesting to note that before the tidal disruption of either star, the binary will be tidally broken apart, so one star's center of mass will be on an unbound trajectory and the other one will be on a bound trajectory. It is possible that the bound star is only partially disrupted (for e.g., $\rp \sim 1.5r_{\rm T,*}$), and in this case, the remnant of the partially disrupted bound star will come back to the pericenter multiple times --- this channel also produces repeating partial TDEs.
For this reason, we suggest that a fraction of the order 1\% the observed TDEs should repeat on a timescale of the order $\rm \sim\!yr$. 

Tight binaries with outer SMAs that do not satisfy eq. (\ref{eq:full_loss_cone_requirement}) are in the empty loss-cone regime for their Hills break-up. In these cases, the inner orbit, even after being tidally hardened to $\ab\lesssim 10R_*$, will again be strongly perturbed before the Hills breakup. This means that the two stars will most likely undergo a collision or merger, instead of Hills breakup \citep{bradnick17_tidal_mergers}. Examples of such mergers may be the G2-like objects \citep{gillessen12_G2}. For instance, the SMA of G2 in our Galactic Center is of the order $0.1\rm\, pc$ \citep{gillessen13_G2_orbit}, which indeed violates our eq. (\ref{eq:full_loss_cone_requirement}), meaning that the original binary's orbit was in the empty loss-cone regime. We note that the stellar merger origin of G2-like objects have previously been considered by many authors \citep[e.g.,][]{prodan15_binary_mergers_GC, stephan16_eKL_nearGC}.


\subsection{Comparison to QPE Models with White Dwarfs}

\label{sec:WD}

\citet{King2020,King2022} argues that WDs on eccentric orbits overflowing their Roche lobe at pericenter can explain the observations of QPEs.   This model has several attractive features:  (1) the need for $10^{4-6} M_\odot$ BHs to tidally strip material off of a WD outside the ISCO is consistent with the low-mass  galaxies in which QPEs have been found \citep{Wevers2022}, although we note there is very large dispersion in galaxy-BH correlations at low masses; (2) given the observed QPE periods and $e \gtrsim 0.9$, it follows that GW driven orbital inspiral can produce mass transfer rates from WDs of order those needed for QPE models --- the same eccentricities lead to viscous times at pericenter $\lesssim$ the orbital period so that in King's models the QPE emission is powered directly by accretion.

There are, however, several issues that we believe argue against WD models for QPEs.   First, as we have argued in \S \ref{sec:disk}, there is relatively good evidence for quiescent accretion disks that are at least as luminous as the QPEs in several systems (e.g., eRO-QPE2 and GSN 69); this argues against viscous times at pericenter less than the orbital period.  More severely, \citet{King2020,King2022} assumes that the mass transfer will be stable in spite of the $R \propto M^{-1/3}$ mass-radius relation of WDs because angular momentum from the disk will couple back to the WD, leading to stability as in stellar binaries.   This is not the case.   The interaction between the star and disk takes place primarily at pericenter, at which the star is moving supersonically relative to the gas.  In this case, the star actually loses angular momentum to the gas, rather than reabsorbing the disk angular momentum (see \S \ref{sec:unstable} for more discussion).\footnote{\citet{King2022} argues for stable mass transfer by analogy to stellar binaries containing mass-transferring WDs.  This analogy does not apply for two reasons.  The first is the high eccentricity $\gg H/r$ just discussed in the text.  The second is the low mass ratio, $M_{*} \ll \Mh$; to elaborate on the latter, we note that  if the orbit were circular, gravitational torques from the star would still probably be too weak to couple strongly to the surrounding disk and reabsorb the disk's angular momentum.  Quantitatively, this can be seen by eq. (\ref{eq:gap}) which requires $H/r \lesssim 4 \times 10^{-3}$ for a $0.3 M_\odot$ WD orbiting a $3 \times 10^5 M_\odot$ BH to open a gap and thus absorb the disk's angular momentum.   By contrast, in King's models $H/r \sim 0.01$ because the accretion rate is $\sim 0.01 \dot M_{\rm Edd}$ and the pericenter of the WD's orbit is $\sim\! 10 \rg$.   Thus eq. (\ref{eq:gap}) is violated (the ratio of the viscous to gravitational torque scales $\sim (H/r)^5$; thus a factor of few violation of eq. (\ref{eq:gap}) is a factor of $\sim 100$ in torque ratio).}   The much higher internal pressure in a WD also implies that the ram pressure due to the disk considered in this paper (\S \ref{sec:sd}) is much less effective at unbinding mass from a WD; it is thus unlikely that the WD's unstable mass transfer can be affected by star-disk interaction.  

It is possible to deliver WDs to an orbit with pericenter radius $\rp\simeq 2R_{*}(M_{\rm BH}/M_{*})^{1/3}$ by Hills breakup of double WD binaries. However, the requirement is that the pre-breakup double WD binary has a very tight inner orbit with $\ab\sim \mr{few}R_{*}$, because otherwise the post-breakup orbital period would be much longer than observed in QPEs. Indeed, the eccentricity of the Hills-capture WD is given by $1-e_{\rm Hills}\simeq \rp \ab/\rTb^2 = 2\times10^{-2} (\ab/R_*)^{-1} (M/10^6M_*)^{-1/3}$, whereas the QPEs require an eccentricity of $1-e_{\rm QPE}\simeq 2\times10^{-2} (P/10\mr{\,hr})^{-2/3} M_{*,0.3}^{-2/3}$ for a WD mass-radius relation $R_{*}=10^{-2}R_\odot(M_{*}/1.4\Msun)^{-1/3}$ \citep{King2022}. The issue for such a tight inner orbit with $\ab\sim \mr{few}R_{*}$ is that the double WD binary only has a GW lifetime of $\sim t_{\rm GWb}\simeq 2\times 10^4\mr{\,yr}\,M_{*,0.3}^{-13/3} (\ab/3R_*)^4$. This lifetime is so short that the binary can only make one or a few orbits around the SMBH if the outer SMA is of the order $a\sim\rm pc$. Thus, the QPE rate can be estimated by
\begin{equation}
    \mc{R}_{\rm QPE, WD}\sim {t_{\rm GWb}\over 10\mr{\,Gyr}} {\rp \over a} \mc{R}_{\rm DWD}\lesssim 10^{-9}\rm\, Gpc^{-3}\,yr^{-1},
\end{equation}
where the factor of ${t_{\rm GW}/10\mr{\,Gyr}}$ is due to the lifetime constraint, and the factor of $\rp/a$ is the fraction of double WD systems with outer-orbit SMA of $a$ that are in the loss cone, $\mc{R}_{\rm DWD}$ is the double WD merger rate contributed by stellar population \textit{near the galactic nuclei}, and to obtain the final upper limit, we have adopted typical values $t_{\rm GWb}\sim 10^4\mr{\,yr}$, $\rp\sim R_\odot$, $a\sim\rm pc$, and $\mc{R}_{\rm DWD}\lesssim 10^5\mr{\,Gpc^{-3}\,yr^{-1}}$ (a conservative limit provided by the type Ia supernova rate). {Hardening of the double WD binary via eccentricity pumping in the inner orbit may also produce $\ab\sim \mr{few}\, R_{*}$, but in nearly all such cases, the hardened double WD binary will merge either on its way to the apocenter of the outer orbit (before coming back to pericenter again) or before angular momentum diffusion can decrease the pericenter distance of the outer binary.  We thus expect the rate of Hills breakup from this channel to be negligible as well.}  We conclude that the rate of delivering WDs to the required orbits is far too small to explain observed QPEs (eq. \ref{eq:obs_QPE_rate}) by Hills breakup of double WD binaries (as also argued by \citealt{Metzger2022}).

\section{Summary and Conclusions}\label{sec:summary}

We have presented general order-of-magnitude considerations regarding the origin of quasi-periodic eruptions (QPEs), a new class of extragalactic transients characterized by $\sim 0.5-8$ hour large amplitude soft X-ray flares with recurrence times of $\sim 2-20$ hrs \citep{miniutti13_GSN069,Miniutti2019,giustini20_RXJ1301,Arcodia2021}.   We argue
that QPEs are powered by unstable mass-transfer between a low-mass $\lesssim 0.5 M_\odot$ star and a supermassive BH (see Fig. \ref{fig:sketch} for a schematic overview).   From this hypothesis, the recurrence time and the condition of Roche-lobe overflow (RLO) for a main-sequence star imply that the stellar orbit is only mildly eccentric with $e \sim 0.5$, that QPEs are not powered by viscous accretion (because the viscous time is much longer than the orbital period), and that GW-driven orbital decay does not set the current lifetime or mass loss rate from the star \citep[as was assumed in some previous works, e.g.,][]{King2022, Metzger2022}. We propose that after an initial phase of unstable mass-transfer, the mass-loss from the star is regulated by ram-pressure and tidal stripping of the star moving inside the accretion disk fed by the star itself (eq. \ref{eq:Mdoteq}). The QPE emission is powered by circularization shocks generated as the stripped stellar debris interacts with the surrounding disk.
Most of our analysis focused on the case of a slowly rotating black hole in which the stellar orbital plane defines the only important angular momentum axis in the system.  The case of a stellar orbit misaligned with respect to the rotation axis of a rapidly rotating black hole is more general but also significantly more complex and requires additional work (see \S \ref{sec:inclined_orbits}).  In addition, the radiation hydrodynamics of stellar mass-loss, star-disk interaction, and QPE production by circularization shocks are quite complex.  Simulations of these processes will be required to make more robust predictions of QPE observables in our model.    

Due to the relatively small emitting area of the shock-heated gas (eq. \ref{eq:hotspot_area}), QPEs have higher effective temperatures than that of the quiescent disk --- the latter is primarily in the difficult to observe EUV band (see Fig. \ref{fig:SED}). Although the accretion efficiency of the quasi-steady disk (a few to 10\%) is likely higher than that of the circularization shocks ($f_{\rm sh}\sim 10^{-2}$ in a typical case where the star has a pericenter radius $\rp\sim 30\rg$), QPEs stand out in the X-ray band because of their hot temperatures --- the peak/quiescent flux ratio may be $\gtrsim 10^2$ near $1\rm\, keV$.  Our model predicts that the time-averaged bolometric luminosity of QPE systems is  dominated by the quiescent disk; this is consistent with the bright very soft quiescent (between QPEs) emission observed in GSN 69 \citep{Miniutti2019} and eRO-QPE2 \citep{Arcodia2021}. {For QPEs with longer periods, the accretion rate will be somewhat lower, which decreases the temperature of the inner disk and hence makes the quiescent emission harder to detect in the X-ray band.}

The quasi-steady disk fed by stellar mass loss has an interesting radial structure, as mass and angular momentum are supplied to the disk near the pericenter of the star's orbit $\rp \sim 30 \rg$ (unlike the AGN or X-ray binary case where mass is fed from much larger radii). The structure at small radii $r<\rp$ is set by a constant accretion rate $\dot{M}$ as in standard disk models, which predicts an emission spectrum $\nu L_\nu \propto \nu^{4/3}$ \citep{Shakura73}. The emission from larger radii $r>\rp$ is given by a disk with a constant angular momentum transport rate $\dot{J}$ (the gas spreads outwards to remove the angular momentum supplied by the stellar debris to the disk); based on this, we predict a spectrum $\nu L_\nu\propto \nu^{12/7}$ in the optical band (Fig. \ref{fig:SED}).  Observed AGN do not have optical spectra consistent with the standard multicolor blackbody prediction of $\nu L_\nu \propto \nu^{4/3}$ \citep{Koratkar1999}.  There is thus reason to be skeptical that $\nu L_\nu\propto \nu^{12/7}$ is realized in nature.   Nonetheless, it would be valuable to carry out high spatial resolution optical observations of QPE sources to see if they have optical spectra steeper than that of standard AGN.   

At the accretion rates $\sim\! 10^{-3} \mspy$ characteristic of QPE sources, the disk is predicted to be radiation pressure dominated \citep{Shakura73}.  Such disks are both thermally and viscously unstable \citep{lightman74_thermal_instability, Piran1978}.   Although there has been progress understanding aspects of these instabilities numerically \citep{Jiang2013,Mishra2022}, the true structure of radiation-dominated accretion disks is not known.   This necessarily leads to uncertainty in some of the predictions of our model.   We do not think that the basic QPE scenario outlined here (Fig. \ref{fig:sketch}) will change significantly as our understanding of radiation dominated disks evolves, but many of the detailed predictions may.  

A second uncertainty in our model is the role of tidal heating as the star's orbit decays towards RLO via GW emission.   The magnitude of the tidal heating depends on the efficiency of non-linear dissipation processes, which are poorly understood (\S \ref{sec:dynamical_tides}).  It is plausible that tidal heating inflates the outer layers of the star initiating RLO and unstable mass transfer at pericenter distances a factor of $\lesssim 2-3$ larger than would occur absent tidal heating.  This would modestly increase the expected rates of eccentric QPEs (and there will be less EMRIs).  The efficiency of tidal heating in a ram pressure confined star is a particularly interesting area for future work: tidal heating is generally most effective in the outer low density layers that are precisely those that are absent once the star is ram-pressure confined by the accretion disk.   

We have also proposed a scenario for the origin of the stellar orbit required in our model, namely a low-mass star with $e \sim 0.5$ and $P \sim 2-20$ hrs around a supermassive BH with $M \sim 10^6 M_\odot$.   Our model (\S \ref{sec:origin}) is based on the tidal breakup of binary stars \citep{hills88_tidal_breakup}. The new ingredients in our model are ones that we believe are generic to the dynamics of stellar binaries around supermassive BHs.    These are: (i) long before the tidal breakup of the binary, the orbital eccentricity of the stellar binary grows to very high values ($\eb\approx 1$) as a result of perturbations by the BH's tidal forces \citep[][]{heggie96_inner_eccentricity, bradnick17_tidal_mergers}; (ii) when the pericenter radius of the inner binary orbit decreases below a critical value that is a few times the stellar radius (see Fig. \ref{fig:rpchaotic}), diffusive growth of the amplitudes of stellar eigenmodes leads to rapid tidal circularization and shrinking of the semi-major axis of the stellar binary to a value $\ab \lesssim 10R_*$.  These very tight binaries are then tidally broken apart when the pericenter of the binary in its orbit around the massive BH become less than about $1\rm\, AU$. One star is ejected as a hyper-velocity star and the other one is captured by the BH with a semi-major axis of the order $100\rm\, AU$.

The bound star left behind after  tidal break-up of the tight stellar binary will eventually undergo RLO or tidal disruption as its orbit continues to evolve due to the combined action of GW orbital decay and angular momentum diffusion due to two-body interactions with other stars and stellar-mass compact objects.   Depending on the eccentricity at the onset of RLO, the resulting mass-transfer can give rise to a repeating partial TDE (a highly eccentric orbit), QPE (mildly eccentric orbit) or EMRI (nearly circular orbit). The relative rates of these phenomena (partial TDEs, QPEs, EMRIs), and the host galaxies in which they reside, depend on the uncertain stellar dynamics in galactic nuclei.

Our current understanding is that a key factor determining the fate of the bound star left after binary disruption is whether the semi-major axis of the star is small or large compared to the ``drain radius'' \citep{alexander04_drain_limit}; the latter is the critical radius below which field star particles are strongly depleted by mutual scatterings.    Interior to the drain radius, stellar relaxation through two-body interactions is less effective and thus GW orbital decay dominates the stellar dynamics. 

We find that for low-mass BHs (roughly $M_{\rm BH}\lesssim10^6\Msun$), the semi-major axis of the Hills-captured star is smaller than the drain radius (eq. \ref{eq:drain_limit_comparison}).  This means that the captured star will undergo gravitational wave inspiral (with angular momentum diffusion playing a minor role) until the start of RLO, leading to QPEs in a significant fraction ($f_{\rm\, QPE}\sim 0.1$) of cases and EMRIs in most of the remaining cases (see Fig. \ref{fig:outcome_fractions}). This connection between the origin of QPEs and EMRIs allows us to predict that the EMRI rate is likely a factor of a few to 10 times higher than the QPE rate. These EMRIs are in roughly circular orbits.  For low-mass main-sequence donors, the mass-transfer will be unstable, producing bright, secularly evolving UV and optical emission. The number density of these bright EMRIs is of the order $10^3\rm\, Gpc^{-3}$, and they may be identified based on their unusually steep optical spectrum of $\nu L_\nu\propto \nu^{12/7}$ and the absence of broad or narrow emission lines.  Some of the  ``EMRIs'' fed by unstable mass transfer from low-mass stars  may in fact produce QPE flares depending on the BH spin and the uncertain outcome of BH-accretion disk alignment due to the \citet{bardeen75_inner_disk_alignment} effect (see \S \ref{sec:inclined_orbits}).   Another class of EMRIs are from higher-mass stars ($M_*\gtrsim 0.5 \Msun$) which undergo stable mass transfer on a much longer GW inspiral timescale. Due to low accretion rates $\dot{M}\sim 10^{-6} M_{\rm BH,6}^{2/3} \mspy$, they are electromagnetically faint. We estimate that the number density of these faint EMRIs is of the order $10^{6}\rm\, Gpc^{-3}\,yr^{-1}$.

For higher-mass BHs (roughly $M_{\rm BH}\gtrsim10^6\Msun$), the semi-major axis of the bound star left behind after binary break-up is larger than the drain radius.  We thus expect the captured star's orbit to undergo significant angular momentum diffusion such that the eccentricity remains high, $1-\eRLO\sim 10^{-2}$, when RLO starts. In these cases, the result is a repeating partial TDE. Possible examples of such events are ASASSN-14ko \citep{payne21_partial_TDE} and eRASSt J0456-20 \citep{liu22_partial_TDE}, which are indeed likely associated with high-mass BHs ($\Mh\sim 10^7\Msun$) based on their host galaxy properties.

Because of the formation of tight ($\lesssim 10 R_*$) stellar binaries prior to the binary's disruption by the massive BH, our model predicts very fast high-velocity stars with asymptotic velocities of $\sim\!2000 M_{\rm BH,6}^{1/6} \rm\, km\,s^{-1}$.  This is consistent with the recently discovered source S5-HVS1 \citep{koposov20_S5-HVS1}.  We predict that such high velocity stars are formed at a time-averaged rate that is a few to 10 higher than the QPE rate, which corresponds to $10^{-6}$--$10^{-5}\rm\,yr^{-1}$ in the Milky Way. Our model also provides a connection between QPEs and other phenomena in galactic nuclei such as double TDEs (where both stars in the inner binary are tidally disrupted) and stellar mergers (where the outcome is likely similar to the class of G2-like objects). More detailed calculations of the velocity distribution of high-velocity stars, and the rates of double TDEs and stellar mergers are left for future work.

\section*{Acknowledgments}
We thank Brian Metzger for useful conversations, and Riccardo Arcodia, Itai Linial, Brian Metzger, and Nick Stone for helpful comments on an initial draft of the paper.  We particularly appreciate Nick's comments that led to \S \ref{sec:dynamical_tides}.  WL was supported by the Lyman Spitzer, Jr. Fellowship at Princeton University. EQ was supported in part by a Simons Investigator grant from the Simons Foundation.  This work benefited from interactions supported by the Gordon and Betty Moore Foundation through grant GBMF5076.

\section*{Data Availability}
The data underlying this article will be shared on reasonable request to the corresponding author.

{\small
\bibliographystyle{mnras}
\bibliography{refs}
}

\appendix

\section{Tidal energy deposition into f-mode}\label{sec:Etidal}
The per-orbit energy deposition into the $\ell=2$ f-mode is given by \citep{press77_tidal_capture}
\begin{equation}
    {\Delta E_{\rm b}\over E_*} = 2(R_*/\rpb)^{6} T_2,
\end{equation}
$E_*=GM_*^2/R_*$, $T_2 = 2\pi^2 \sum_n Q_{n2}^2 \sum_{m=-2}^{2}K_{n2m}^2$, $Q_{n2}\approx 0.5$ is the spatial overlap integral for the f-mode ($n=0$), and $K_{n\ell m}$ is the temporal overlap integral between the mode and the time-dependent tidal potential of the companion. The factor of 2 in $\Delta E_{\rm b}$ comes from the assumption that two identical stars both get tidally excited. For our purpose here, the dominant contribution comes from the prograde mode $m=-2$ and we have
\begin{equation}
\begin{split}
    K_{02-2} = {2\sqrt{2}\over \sqrt{15}}& {\omega_*\over \Omega_{\rm peri, b}} \lrb{1 + {\eps\over 2}} z^{3/2} \mr{e}^{-2(1+\eps/5)z/3}\\
    &\times \lrsb{1 - {\pi^{1/2}\over 4}\lrb{1 + {7\eps\over 2}} z^{-1/2}},
\end{split}
\end{equation}
where $\Omega_{\rm peri,b} = \sqrt{2GM_*(1+\eb)/\rpb^3}$ is the pericenter angular frequency, $\eps=(1-\eb)/(1+\eb)$, and $z=2\omega_{\rm f}/\Omega_{\rm peri,b}$.
The above analytical expression is based on direct integration of the temporal overlap function using the saddle point method (Lu et al., in prep), which is an extension of the parabolic ($\eb=1$) case worked out by \citet{lai97_tidal_energy}.

\section{Stellar Mass-Loss Generated by a Strong Shock} \label{sec:outer_layers}
The structure of the outer layers of the star ($R\approx R_*$) can be obtained from the condition of hydrostatic equilibrium combined with a polytropic equation of state $p=K\rho^\gamma$ ($K$ being the entropy constant, and $\gamma$ being the polytropic index).  The  outer layers of the star are not well-described by a polytropic model once they are subject to a strong shock; indeed, the outer parts of the star are shocked to much higher entropy than the inner parts.   However, we are interested here in determining the properties of the star at depths where the shock does not have a significant effect; this is determined by the pre-shock structure of the star, which can be adequately modeled as polytrope with $\gamma \simeq 5/3$ for low-mass stars.  Hydrostatic equilibrium is thus given by
\begin{equation}
    {\d p\over \d R} = \frac{5 K}{3} \rho^{2/3} {\d \rho\over \d R} \approx -{GM_*\over R_*^2} \rho,
\end{equation}
which can be integrated to give the sound speed profile
\begin{equation}
    \cs^2(\Delta R) = K\rho^{2/3} = \frac{2}{5} {GM_*\over R_*^2} \Delta R,
\end{equation}
where $\Delta R=R_*-R$ is the exterior radius.
Ignoring a factor of order unity, the pressure profile in the outer layers is roughly given by
\begin{equation}
    {p(\Delta R)\over \bar{p}} \sim (\Delta R/R_*)^{5/2},
\end{equation}
where $\bar{p} = GM_*^2/(4\pi R_*^4)$ is the average pressure of the star. The density and temperature profiles have the following power-law scalings $\rho\propto \Delta R^{3/2}$ and $T\propto \Delta R$. Since the pressure profile is directly related to the exterior mass by $p(\Delta R)/\bar{p} = M_{\rm ex}/M_*$ (eq. \ref{eq:PM}), we obtain
\begin{equation}
    {\Delta R(M_{\rm ex})\over R_*} \sim \lrb{M_{\rm ex}/ M_*}^{2/5}.
\end{equation}

When the star is placed under an external ram pressure $p_{\rm ram}\ll \bar{p}$, the outer layers will be shock-heated down to a depth of $\Delta R(p_{\rm ram}) \sim R_* (p_{\rm ram}/\bar{p})^{2/5}$.
When the star reaches near pericenter where an unperturbed star would marginally fill up its Roche lobe, the shock-heated outer layers can expand to a distance $\Delta R$ beyond the Roche lobe. This leads to a mass loss rate of the order
\begin{equation}\label{eq:L1_mass_loss_rate}
    \dot{M}_* \sim A_{\rm st} \rho(\Delta R) \cs(\Delta R),
\end{equation}
where $A_{\rm st}\sim R_*\Delta R$ is the cross-sectional area of the nozzle near the L1 point, $\rho(\Delta R) \sim M_{\rm ex}(\Delta R)/(4\pi R_*^2 \Delta R)$ is the density of the overflowing layer, and $\cs\sim \sqrt{GM_*\Delta R/R_*^2}$ is the sound speed in that layer. There may be additional mass loss from the L2 nozzle (if $\Delta R/R_*$ significantly exceeds $(M_*/M_{\rm BH})^{1/3}$), but that only gives an order-unity correction to the mass loss rate. 

Near pericenter, the orbital radius $r$ evolves with the true anomaly $\Phi$ as
\begin{equation}
    {r\over \rp} \approx 1 + {2e\over 1+e} \lrsb{\tan(\Phi/2)}^2,
\end{equation}
so the star only fills up the Roche lobe within a maximum true anomaly that is given by
\begin{equation}
    \Phi_{\rm max} = \mr{atan}\lrb{\sqrt{2(1+e)\Delta R\over e R_*}} \simeq \lrsb{2(1+e)\Delta R\over e R_*}^{1/2},
\end{equation}
where the second expression is valid when $\Phi_{\rm max}\ll 1\rm\, rad$. For $\Delta R/R_*\sim 10^{-2}$ (corresponding to $p_{\rm ram}/\bar{p}\sim 10^{-5}$) and $e=0.5$, we have $\Phi_{\rm max} \simeq 0.2\rm\, rad$. Thus, each episode of Roche-lobe overflow near pericenter only lasts for a time
\begin{equation}
    t_{\rm RLO} \sim {2\Phi_{\rm max} \over \Omega_{\rm peri}} \sim \lrb{{8\Delta R\over e R_*} {\rp^3\over GM_{\rm BH}}}^{1/2},
\end{equation}
where $\Omega_{\rm peri} = (1+e)^{1/2}\OmgK^{-1}(\rp)$ is the angular frequency of the star near pericenter. For a pericenter radius of $\rp = 2R_*(\Mh/M_*)^{1/3}$ (eq. \ref{eq:RLO_radius}), we obtain the mass loss per orbit
\begin{equation}\label{eq:dMstar_per_orbit}
    \Delta M_* = \dot{M}_* t_{\rm RLO} \sim {2M_{\rm ex} \Delta R\over \pi e^{1/2} R} \sim \lrb{p_{\rm ram}\over \bar{p}}^{7/5}M_*,
\end{equation}
where the second expression applies for a modest eccentricity that is not extremely close to zero (we are only aiming at obtaining an order-of-magnitude estimate of the per-orbit mass loss).  We note that, for general $\gamma$, eq. (\ref{eq:dMstar_per_orbit}) would become $\Delta M_*/M_* \simeq (p_{\rm ram}/\bar p)^{2 - 1/\gamma}$ which is not very different from eq. (\ref{eq:dMstar_per_orbit}) for reasonable variation in $\gamma$.

The outer layers of the star at depths much less than $\Delta R(p_{\rm ram})$ have higher post-shock sound speeds and hence will expand farther from the star. It can be easily shown that a layer of mass $\sim (p_{\rm ram}/\bar{p})^{5/3} M_*$ will be able to expand to $\sim 2R_*$; this contributes a smaller per-orbit mass loss than in eq. (\ref{eq:dMstar_per_orbit}). However, if the pericenter of the orbit is located at a few times $\rRLO$ (in the case where tidal heating inflates the star and leads to an earlier onset of RLO, see \S \ref{sec:dynamical_tides}), then the mass loss per orbit would be given by
\begin{equation}\label{eq:dMstar_per_orbit_alternative}
    \Delta M_* \sim (p_{\rm ram}/\bar{p})^{5/3} M_*.
\end{equation}
The above expression (instead of eq. \ref{eq:dMstar_per_orbit}) will slightly decrease the equilibrium mass-loss rate in eq. (\ref{eq:Mdoteq}) to $\dot{M}_{\rm eq}\simeq 9\times 10^{-4}\mspy\, M_{\rm BH,6}^{5/13} M_{*,0.5}^{9/13} r_{30}^{5/26} \alpha_{0.1}^{-5/13} P_{10}^{-3/13}$. We note that the estimates in eqs. (\ref{eq:dMstar_per_orbit}, \ref{eq:dMstar_per_orbit_alternative}) are based on an unperturbed star, whereas, in reality, the star's outer layers will be subjected to multiple episodes of shock interactions, as long as $\Delta M_* < M_{\rm ex}$ (i.e., the per-orbit mass loss is less than the shock-heated mass). Detailed numerical simulations are needed to understand the long-term evolution of the star. However, we expect that our overall conclusions are relatively insensitive to the details on the stellar response under periodic ram-pressure perturbations.


\label{lastpage}
\end{document}